\documentclass[prstab,amsmath,amssymb,aps,showkeys,showpacs,a4paper,reprint]{revtex4-1}
\pdfoutput=1
\usepackage{ifpdf}
\usepackage{graphicx,amssymb}
\usepackage{graphicx}
\usepackage{dcolumn}
\usepackage{bm}
\usepackage{rotating}
\makeatletter
\usepackage{amsmath}
\usepackage{array}
\usepackage{booktabs}
\usepackage{dcolumn}
\usepackage{subfigure}
\newcolumntype{d}[1]{D{.}{.}{#1}}
\graphicspath{{figs/}{../../../ENC/Habil/figs/}}
\makeatletter
\newcommand{\todo}[1]{}%
\makeatother

\newcommand{\ud}{\ensuremath{\mathrm{d}}}%
\newcommand{\rref}{R_{Ref}}%
\renewcommand{\imath}{\mathtt{i}}%
\newcommand{\cplx}[1]{{\mathbf {#1}}}%
\newcommand{\conj}[1]{\overline{\cplx{#1}}}%
\newcommand{\zsarg}{\frac{\cplx{z}}{\rref}}%
\newcommand{\zs}{\left(\zsarg\right)}%
\newcommand{\re}{\operatorname{Re}}%
\newcommand{\im}{\operatorname{Im}}%
\newcommand{\kroneckerdelta}{\delta}%
\newcommand{\uv}[1]{%
  \vec{i}_{#1}%
}%
\makeatletter
\newcommand{\leftrightarrowfill}{%
  $\m@th \smash -\mkern -7mu\cleaders%
  \hbox{$\mkern -2mu\smash -\mkern -2mu$}%
  \hfill \mkern -7mu\mathord \leftrightarrow $%
}%
\makeatother
\newcommand{\pt}{\partial}%
\renewcommand{\div}{\nabla\cdot}%
\newcommand{\grad}{\nabla}%
\let\vph=\varphi%
\let\vth=\vartheta%
\let\eps=\varepsilon%
\let\no=\nonumber%
\newcommand{\rc}{R_C}%
\newcommand{\rr}{\rref}%
\newcommand{\ordersymbol}{\operatorname{\mathcal{O}}}%
\newcommand{\order}[2]{\ordersymbol\left({#1}^{#2}\right)}%
\newcommand{\ordepsq}{\order{\epsilon}{2}}%
\newcommand{\cz}{\cplx{z}}%
\newcommand{\bPh}{{\bar{\Psi}}}%
\newcommand{\vbT}{\vec{\bar{T}}}%
\newcommand{\abs}[1]{{\left|{#1}\right|}}%
\newcommand{\be}{\bar{e}}%
\newcommand{\bae}{\bar{\eta}}%
\newcommand{\baps}{\bar{\psi}}%
\newcommand{\beps}{\bar{\epsilon}}%
\newcommand{\bh}{\bar{h}}%
\newcommand{\mat}[1]{\mathit{#1}}%
\newcommand{\upmat}{\mathit{\mathcal{L}}}%
\newcommand{\fr}{\frac}
\newcommand{\bE}{\bar{E}}

\newcommand{\bfz}{\mathbf{z}}
\newcommand{\bfw}{\mathbf{w}}
\newcommand{\bfC}{\mathbf{C}}
\newcommand{\bfE}{\mathbf{E}}
\newcommand{\bfB}{\mathbf{B}}
\newcommand{\bfPhi}{\mathbf{\Phi}}
\newcommand{\bfXi}{\mathbf{\Xi}}

\begin{document}

\title{Cylindrical Circular and Elliptical, Toroidal Circular and Elliptical Multipoles\\ Fields, Potentials
  and their Measurement for Accelerator Magnets}
\date{\today}
\author{Pierre Schnizer}
\email{p.schnizer@gsi.de}
\author{Egbert Fischer}
\email{e.fischer@gsi.de}
\affiliation{GSI Helmholtzzentrum f\"ur Schwerionenforschung mbH,
 Planckstra\ss{}e 1, 64291 Darmstadt, Germany}

\author{Bernhard Schnizer}
\email{bernhard.schnizer@tugraz.at}
\affiliation{Institut f\"ur theoretische Physik - Computational Physics, Technische Universit\"at
 Graz, Petersgasse 16, 8010 Graz, Austria}


\begin{abstract}
  Recent progress in particle accelerator tracking has shown that the 
  field representation is one of the major limits of the prediction accuracy especially
  for machines, whose aperture is fully filled by the beam and thus higher the artefacts
  created by higher order   modes have to be thoroughly understood.
  
  The standard tool for field presentation today are cylindrical circular multipoles
  due to their straight forward correspondence to the Cartesian coordinates.
  In this paper we extend the standard approach to other coordinate systems, show
  how these can be measured next to their realisation in measuring the SIS100 Magnets
  for the FAIR project.
\end{abstract}

\maketitle



\section{Introduction}

Studying the performance of an existing or to be built accelerator
requires solid knowledge on the magnetic field quality so that the
expected beam behaviour can be calculated. Nowadays the field of each
type of magnet can be calculated by numerical codes. But analytic
expressions of the portion of the static or quasi-static magnetic
field in the gap are needed for beam dynamics calculations. In that
part of the gap which is free from conductors or charges the field is
a potential field. Therefore expansions of the potential are used
comprising particular solutions of the potential equation to describe
this field. Their coefficients are obtained by fitting the expansions
to the numerical field values produced by the code. Such particular
solutions are obtained by solving the potential equation by separation
and are called multipoles.

All textbooks related to accelerator physics have been using local
Cartesian or cylindric circular coordinate systems to describe the
field in long magnets. (Transverse deviations of the particle from the
ideal orbit are followed by the Frenet Serret coordinates.)  While
this approach is general and has proven to be applicable by
experience and manageable with the computing power typically available
in the last decade, 
producing the circular expansion coefficients has shown to
be troublesome for practical accelerator magnets, in particular for
iron dominated ones, where the height of the gap is considerably smaller
than the aperture width. Here the standard approach calculating the
multipoles over a circular boundary will not work as $\Delta \Phi= 0$
is not defined everywhere if the circle is chosen with a radius equal
to half the width of the aperture. Similarly data obtained from the
boundary of a circle with a radius equal to half the height will
represent the field only slightly beyond the measurement radius. 

During the R\&D phase of the heavy ion synchrotron SIS100 these
problems had to be tackled as the beam uses considerable area of the
elliptic vacuum chamber. This led to the development of elliptic
cylindric multipoles \cite{pbep,pbep:mt22,pbep:NIMA}.

During the R\&D phase the design of the dipole magnets, originally
straight 2 T, 4 T/s, 2.65 m long magnets, were changed to curved ones
with a radius of curvature of 52.625 m. As the advanced beam dynamics
studies required a set of reliable harmonics, a concise solution was
required to be able to develop a measurement concept and to extract
data for beam dynamics use. The appropriate multipoles were derived
using local toroidal coordinates and the technique of
R-separation, as these give simpler solutions, as the global toroidal ones 
(see e.g. \cite{MS,global-toroidal-brouwer-pac13}),
and are easier  to interpret \cite{pbep:compel,pbep:mt22}.
 The solution was then further extended to elliptic
toroidal multipoles \cite{pbep:igte2010}.

All these developments were driven by the limitations we found in the
standard tool, the cylindric circular multipoles, and were developed for designing the
SIS100 synchrotron, measuring the dipole magnets and predicting the
performance of the machine.

This paper is split into the following parts:
\begin{enumerate}
\item 
  First the mathematics of the cylindrical circular multipoles is
  recalled.
\item 
  The different new coordinate systems are described in a way,
  which always clearly shows their relation to the preceding system 
  next to the properites of the obtained solutions.
\item 
  Finally their application is illustrated showing how these
  multipoles can be measured.
\end{enumerate}

While these developments have been made for SIS100, the solutions obtained are 
applicable to any problem formulated by the potential equation in the 
geometries given above.



\section{Theory}

\subsection{Plane Circular Multipoles}
\label{sec:circular-multipoles}

%

The circular multipoles are the common workhorse for representing the two-dimensional field in the
gap of long straight accelerator magnets. Here just the most important formulas are listed. Details
may be found in the papers of  \cite{jain98:_magnet} and \cite{cas:magnets:wolski}.

Circular multipoles are particular regular solutions of the potential equation 
\begin{equation}
\Delta\Phi_r \ = \ 0 
\label{PotEq}
\end{equation}
in Cartesian coordinates $(x, y)$ or in polar coordinates $\rho, \theta$ with $x = \rho \ \cos\theta, \ y = \rho \ \sin\theta:$
\begin{eqnarray}
\label{PotEqCart}
\Delta\Phi_r \ = \ \left( \fr{\pt^2}{\pt x^2} \ + \  \fr{\pt^2}{\pt y^2}  \right) \Phi_r  &=& 0, \\
\label{PotEqPolar}
\Delta\Phi_r \ = \ \left( \fr{\pt^2}{\pt r^2} \ + \ \fr{1}{r} \fr{\pt}{\pt r} \ + \fr{\pt^2}{\pt \theta^2}  \right) \Phi_r  &=& 0 .
\end{eqnarray}
Any non-negative integer power 
\begin{eqnarray}
\bfC_m \ (\bfz /\rref)^m &=& \bfC_m \ [(x + \imath y)/\rref]^m \\
\nonumber
&=& \bfC_m \  (r/\rref)^m e^{\imath m \theta}, \,\ m = 0, 1, 2, ...
\end{eqnarray}
is such a complex
regular solution of Eqs.~\eqref{PotEqCart} or \eqref{PotEqPolar}. The reference radius $\rref$ is inserted to render all
the solutions dimensionless and of similar magnitude.  The complex constants $\bfC_m $ determine the magnitude and
the direction of each multipole. In practice the  irrotational and source-free magnetic induction of a single ideal multipole 
may be written in a concise complex representation as:
\begin{eqnarray}
\bfB(\bfz)  &:=& B_y (x,y) + \imath \ B_x (x,y) \\
\nonumber
 &=& \ \bfC_m \ (\frac \bfz \rref)^{m - 1} = \ \bfC_m  \ \left(\frac r \rref\right)^{m-1} \ e^{\imath(m - 1)\theta}.
\label{CMm}
\end{eqnarray}
The induction in a real long magnet is a superposition of such multipoles:
\begin{equation}
\bfB^C (\bfz) \ := B^C_{y} (x,y) + \imath \ B^C_{x} (x,y) =  \ \sum_{m=1}^M \bfC_m \ (\bfz /\rref)^{m - 1}.
\label{CMAll}
\end{equation}
Here the European convention is adopted: m = 1 gives a dipole field, m = 2 a quadrupole field, a.s.o.
$M$ is the number of multipoles used; in theory $M = \infty$, in practice $M$ is about 20.

The two equations above define complex functions having a complex potential:
\begin{equation}
\bfB^C (\bfz) \ = \ -  \rref\ \ d\bfPhi^C/d\bfz. 
\end{equation}
A simple integration gives:
\begin{equation}
\bfPhi^C (\bfz) = -  \ \sum_{m=1}^M \fr{1}{m} \ \bfC_m \ (\bfz /\rref)^{m} = - \  \sum_{m=1}^M  \bfC_m \ \mathbf{\Phi}_m^C (\mathbf{z}).
\label{PCMAll}
\end{equation}
A zero value has been assumed for the integration constant; so $\bfPhi^C (\bfz=0) = 0$. 

\subsubsection{Normal circular multipoles}

Assuming real values of the constants $ \bfC_m = B_m$, then taking the imaginary part of the resulting potential $\bfPhi^C (\bfz)$, Eq.~\eqref{PCMAll}
gives the real potential $\Phi^{Cn} (x,y)$.  
The real components of the magnetic induction can be computed from this real potential:
\begin{eqnarray}
\label{BCn}
(B^{Cn}_{x}, B^{Cn}_{y}) &=& - \rref \ \text{grad} \Phi^{Cn} (x,y) 
\\
\nonumber
&=& \ - \rref\ \left( \fr{\pt}{\pt x}, \fr{\pt}{\pt y} \right) \Phi^{Cn} (x,y). \\
\label{PhCn}
 \Phi^{Cn} &=&  - \ \sum_{m=1}^M B_m  \Phi^{Cn}_m \ \\
\nonumber
&=& \ - \ \sum_{m=1}^M B_m \fr{1}{m} \im \left( \fr{\bfz}{\rref} \right)^{m}  .
\end{eqnarray}
The first terms are listed in Table~\ref{tab:circular-cylindric-multipoles}.
\begin{table}
  \caption{Cylindric circular multipoles,  first terms of the
    potential and the basis functions.}
  \label{tab:circular-cylindric-multipoles}
  \centering
  {
    \begin{ruledtabular}
      \begin{tabular}[c]{rccc}
      & $\Phi_r(x,y)$ & $B_x(x,y)$& $B_y(x,y)$ \\
      \hline
      \multicolumn{4}{l}{normal}\\
      
 1  & $ - \frac{y}{\rref} $  & $ 0 $ & $ 1 $\\ 

 2  & $ -  \frac{x y}{\rref^{2}} $  & $ \frac{y}{\rref} $ & $ \frac{x}{\rref} $\\ 

 3  & $ \frac{y \left(- 3 x^{2} + y^{2}\right)}{3 \rref^{3}} $  & $ 2 \frac{x y}{\rref^{2}} $ & $ \frac{x^{2} - y^{2}}{\rref^{2}} $\\ 

 4  & $  \frac{x y \left(- x^{2} + y^{2}\right)}{\rref^{4}} $  & $ \frac{y \left(3 x^{2} - y^{2}\right)}{\rref^{3}} $ & $ \frac{x \left(x^{2} - 3 y^{2}\right)}{\rref^{3}} $\\ 

      \hline
      \multicolumn{4}{l}{skew}\\
      
 1  & $ - \frac{x}{\rref} $  & $ 1 $ & $ 0 $\\ 

 2  & $ \frac{- x^{2} + y^{2}}{2 \rref^{2}} $  & $ \frac{x}{\rref} $ & $ - \frac{y}{\rref} $\\ 

 3  & $ \frac{x \left(- x^{2} + 3 y^{2}\right)}{3 \rref^{3}} $  & $ \frac{x^{2} - y^{2}}{\rref^{2}} $ & $ - 2 \frac{x y}{\rref^{2}} $\\ 

 4  & $ \frac{- x^{4} + 6 x^{2} y^{2} - y^{4}}{4 \rref^{4}} $  & $ \frac{x \left(x^{2} - 3 y^{2}\right)}{\rref^{3}} $ & $ \frac{y \left(- 3 x^{2} + y^{2}\right)}{\rref^{3}} $\\ 
%

    \end{tabular}
  \end{ruledtabular}
  }
\end{table}
A zero value has been assumed for the integration constant; so $\ \Phi^{Cn}(\bfz=0) = 0$. 
The normal multipole fields are a vertical
induction $B_{y}$ for  m = 1:  a quadrupole with pole faces normal to the coordinate axes at x = 0, y = 0 
respectively, for m = 2; a.s.o.  The same field expressions as in Eq.~\eqref{BCn} are also found by taking real and imaginary parts of the
complex field representation $\bfB^C (\bfz)$, Eq.~\eqref{CMAll}.

\subsubsection{Skew circular multipoles}

Assuming purely imaginary values of the constants $ \bfC_m = i A_m$,  then taking the imaginary part of the resulting potential $\bfPhi^C (\bfz)$, Eq.~\eqref{PCMAll},
gives the real potential $\Phi^{Cs}(x,y)$.  
The real components of the magnetic induction can be computed from this real potential:
\begin{eqnarray}
\label{BCs}
(B^{Cs}_x, B^{Cs}_y) &=& - \rref \ \text{grad} \Phi^{Cs} (x,y) \\
\nonumber
&=& \ - \rref\ \left( \fr{\pt}{\pt x}, \fr{\pt}{\pt y} \right) \Phi^{Cs} (x,y);  \\
\label{PhCs} 
 \Phi^{Cs} (x,y) &=& - \ \sum_{m=1}^M A_m  \Phi^{Cs}_m \\ 
\nonumber
& = &  -  \sum_{m=1}^M A_m \fr{1}{m}  \text{Re} \left( \fr{\bfz}{\rref} \right)^{m}   .
\end{eqnarray}
A zero value has been assumed for the integration constant; so $\ \Phi^{Cs}(\bfz=0) = 0$. 

So one gets skew multipoles. This is a purely horizontal magnetic
induction $B_x$ for m = 1;  a quadrupole with pole faces normal to the bisectors of the coordinate axes for m = 2; a.s.o.
The same field expressions as in Eq.~\eqref{BCs} are also found by taking real and imaginary parts of the
complex field representation $\bfB^C(\bfz)$, Eq.~\eqref{CMAll}.

\subsection{Elliptic multipoles}
\label{sec:elliptic-multipoles}

\subsubsection{Elliptic coordinates}

Elliptic coordinates are superior to circular coordinates in gaps of elliptic cross section,
as typically used as beam aperture when iron dominated magnets are used for guiding the beam.
An ellipse as reference curve covers a larger area than an inscribed circle. The reference ellipse
is defined by its semi-axes $a$ and $b$ giving the eccentricity $e$. 
Plane elliptic coordinates  $\eta, \psi$ may be introduced by a conformal mapping:
\begin{eqnarray}
\bfz := x + \imath y = e \ \cosh \bfw, &\ \ & \bfw := \eta + \imath \psi; 
\label{eq:elliptic-coordinates}\label{EllCord}\label{eq:coor-ellipse-eccentricity} \\
\bfw =  \mbox{Arcosh}(\bfz/e), &\ \ & \eta_0 =  \mbox{ArTanh}(b/a)
\end{eqnarray}
$\eta_0$ is the value of the quasi-radial variable $\eta$ corresponding to the reference ellipse.
Taking the real and the imaginary part of Eq.~\eqref{EllCord} gives a real vector ${\bf r}= (x, y)$. Computing the
tangent vectors ${\bf r}_\eta$ and ${\bf r}_\psi$, normalising and generalising them gives formulas 
for transforming components of the same vector $\mathbf{a}$ between the Cartesian and the elliptic system:
\begin{eqnarray}
a_\eta &=& \quad \sinh\eta \ \cos\psi \ a_x/h_t  \ + \  \cosh\eta \ \sin\psi \ a_y/h_t , \label{Tftaxae}\\ 
a_\psi &=& - \ \cosh\eta \ \sin\psi \ a_x/h_t  \ + \  \sinh\eta \ \cos\psi \ a_y/h_t ; \no\\
h_t      &=&  e \  \sqrt{\cosh^2 \eta \ \sin^2 \psi \ + \ \sinh^2 \eta \ \cos^2 \psi} = \\
&=&  e \ \sqrt{\cosh^2 \eta - \cos^2 \psi} \ = \ e \ \sqrt{\sinh^2 \eta + \sin^2 \psi} \no
\end{eqnarray}
The arc element is: 
\begin{equation}
ds^2 = dx^2 + dy^2 = h_t^2 \left( d\eta^2 + d\psi^2 \right).
\end{equation}

\subsubsection{Elliptic multipole field expansions for Cartesian components depending on elliptic coordinates}

Solutions of the potential equation, so multipoles in elliptic coordinates were introduced and discussed 
at length in \cite{pbep:NIMA}.
The first basis terms 
are given in Table~\ref{tab:field-descriptions-cylindric-elliptic-multipoles}.
\begin{table}[b]
  \caption{  
    The basis functions of the cylindric elliptic multipoles.}
  \label{tab:field-descriptions-cylindric-elliptic-multipoles}
  \centering
  {
      \begin{ruledtabular}
    \begin{tabular}[c]{rccc}
      & $\Phi^e_r$ & $B_x(x,y)$& $B_y(x,y)$ \\
      \hline
      \multicolumn{3}{l}{normal}\\
       1 & $-\frac{\psi }{2}                            $ & $ 0                             $ & $\frac{1}{2}                   $ \\
       2 & $-\cosh (\eta ) \sin (\psi )                 $ & $ \sin (\psi ) \sinh (\eta )    $ & $\cos (\psi ) \cosh (\eta )    $ \\
       3 & $-\frac{1}{2} \cosh (2 \eta ) \sin (2 \psi ) $ & $ \sin (2 \psi ) \sinh (2 \eta )$ & $\cos (2 \psi ) \cosh (2 \eta )$ \\
       4 & $-\frac{1}{3} \cosh (3 \eta ) \sin (3 \psi ) $ & $ \sin (3 \psi ) \sinh (3 \eta )$ & $\cos (3 \psi ) \cosh (3 \eta )$ \\
       \hline
      \multicolumn{3}{l}{skew}\\
      1 & $-\frac{\eta }{2}                             $ & $ \frac{1}{2}                   $  & $0                              $ \\
      2 & $-            \sinh (  \eta ) \cos (\psi )    $ & $ \cos (\psi ) \cosh (\eta )    $  & $-\sin (\psi ) \sinh (\eta )    $ \\
      3 & $-\frac{1}{2} \sinh (2 \eta ) \cos (2 \psi )  $ & $ \cos (2 \psi ) \cosh (2 \eta )$  & $-\sin (2 \psi ) \sinh (2 \eta )$ \\
      4 & $-\frac{1}{3} \sinh (3 \eta ) \cos (3 \psi )  $ & $ \cos (3 \psi ) \cosh (3 \eta )$  & $-\sin (3 \psi ) \sinh (3 \eta )$ \\
    \end{tabular}
  \end{ruledtabular}
  }
\end{table}
 Here only some
important formulas are quoted from this source and some new results will be given.
Analogous to Eq.~\eqref{CMAll} we define a complex field representation depending on the elliptic 
coordinates $\eta, \psi$ through $\bfw =  \eta + \imath \psi$ as:
\begin{eqnarray}
\label{CEAll1}
\bfB^{Ce} (\bfw)  & =&  B^{Ce}_y (\eta, \psi) + \imath B^{Ce}_x (\eta, \psi) \\
\nonumber
&:=& \fr{\bfE_1}{2} \ + \ \sum_{k=2}^M \bfE_k \ \frac{\cosh \big((k - 1) ( \eta + \imath \psi) \big)}{\cosh((k - 1) \eta_0)}  \\
\label{CEAll2} 
& :=& - \ d \bfXi^{Ce} / d \bfw    \\
\nonumber
&=& \sum_{k=1}^M \bfE_k \  \frac{1}{1 + \delta_{k1}} \ \frac{\cosh ((k - 1) \bfw )}{\cosh((k - 1) \eta_0)} 
\end{eqnarray}
$\cosh\big(= \ \bfC_m  (\eta + \imath \psi) \big)$ is a regular solution of the potential equation in elliptic coordinates:
\begin{equation}
\Delta\Phi_r^e = \fr{1}{h^2_t} \left[ \fr{\pt^2}{\pt\eta^2} \ + \ \fr{\pt^2}{\pt\psi^2}  \right]\Phi_r^e = 0.
\end{equation}
\begin{widetext}
To relate the elliptic multipoles to the circular ones we use the following formula (\cite{GR}, Eq.~1.331.4)
\begin{eqnarray}
\cosh(\nu \bfw ) &=&  2^{(\nu-1)} \cosh^\nu \bfw + \sum\limits_{\mu=1}^{[\nu/2]} (-1)^\mu \ \fr{\nu}{\mu} 
\left( \begin{array}{c}  \nu - \mu - 1   \\  \mu - 1 \end{array} \right) 2^{(\nu - 2\mu -1)} \cosh^{\nu - 2\mu}\bfw  \label{MultCosh} \\
&=& 2^{(\nu-1)} \left( \fr{\rref}{e}  \fr{\bfz}{\rref}  \right)^\nu 
 +  \sum\limits_{\mu=1}^{[\nu/2]}  (-1)^\mu \ \fr{\nu}{\mu} 
\left( \begin{array}{c}  \nu - \mu - 1   \\  \mu - 1 \end{array} \right)  2^{(\nu - 2\mu -1)} \left( \fr{\rref}{e}  \fr{\bfz}{\rref}  \right)^{\nu - 2\mu} 
\end{eqnarray}  
\end{widetext}
$[\nu/2]$ is the largest integer equal to or just below $\nu/2$. The equation above
shows that $\bfB^{Ce} (\bfw)$ is again a linear superposition of plane circular multipoles Eq.~\eqref{CMm}.
But the coefficients of this new series may be computed from data given along the reference ellipse. These data
contain more accurate information on higher multipoles. Practical applications show that this new series converges faster 
and less aleatory \cite{pbep:NIMA}.  Even when working with circular multipoles in an elliptic aperture
it is advantageous to use expansion coefficients $\bfC_m$ computed from the elliptic coefficients $\bfE_k$:
\begin{equation}
\bfC_m   \ = \ \left( \frac{\rref}{e} \right)^{m-1} \ \sum_{k=1}^M \bfE_k \ \frac{(1 + \delta_{k1})^{-1}}{\cosh((k-1) \eta_0) } t_{k-1,m-1}.
\end{equation}
The elements of the real transformation matrix, $t_{ms}$, have been derived in  \cite{pbep:NIMA} by a complex integration and 
Cauchy's residue theorem. An equivalent simpler formula  found by rewriting Eq.~\eqref{MultCosh} as
\begin{equation}
\cosh(k \bfw ) \ =  \ \sum_{\nu=0}^k t_{k\nu} \ \cosh^{\nu} w , \quad k = 0, 1, ..., n. 
\end{equation}
is given here:
\begin{equation}
t_{k\nu} \!=\! \left\{
\!\!\!
\begin{array}{rcl}
&   0        & \text{if} \quad k + \nu =  \text{odd}  \,  \vee \, k < \nu,  \\
& 1          & \text{if} \quad k = \nu = 0,  \\
& 2^{k - 1}    & \text{if} \quad  k = \nu \ge 1, \\
& t^{k>\nu}_{k\nu} & \text{if} \quad k - \nu =  \text{even} \, \wedge \,  k > \nu. 
\end{array} \right.
\end{equation}
\begin{equation}
t^{k>\nu}_{k\nu} = \frac{k\, 2^\nu\, (-1)^{(k + 3 \nu)/2 }}{k-\nu} \binom{\frac{k+\nu}{2}-1}{\frac{k-\nu}{2}-1}
\end{equation}
$(t_{k\nu})$ is a lower triangular matrix with a nonzero main diagonal.
An even more concise formula due to \cite{elliptic_to_circular:vassili}
\begin{equation}
t_{k\nu} = \mbox{Coefficient}[T_k(w),w^\nu]
\end{equation}
uses Chebyshev polynomials defined by:
\begin{eqnarray}
T_k(w) &=& \cos (k \ \arccos (w)),  \quad - 1 \leq w \leq 1;\\
 &=& \cosh(k \ \mbox{Arcosh}(w)), \quad 1 \leq  w < \infty.
\end{eqnarray}

The complex regular function given in Eq.~\eqref{CEAll2} may be integrated w.r.t. $\bfw$ to give the 
auxiliary function $\bfXi^{Ce} (\bfw)$ belonging to Eq.~\eqref{CEAll2}:
\begin{equation}
\bfXi^{Ce} (\bfw) \ = \  - \ \fr{\bfE_1}{2} \bfw \ -  \ \sum_{k=2}^M \fr{1}{k-1} \ \bfE_k \ \frac{\sinh ((k-1) \bfw )}{\cosh((k-1) \eta_0)}.
\label{CEXAll}
\end{equation}
A zero value has been assumed for the integration constant; so $\bfXi^{Ce} (\bfw = 0)  = 0$. 
$\bfXi^{Ce} (\bfw)$ is not a single-valued potential. One may derive the Cartesian components of the magnetic induction by the derivatives 
given in Eqs.~(\eqref{BnfrXin}) and (\eqref{BsfrXis}). But it is not the operator grad of the elliptic coordinates,
which transforms $\bfXi^{Ce} (\bfw)$ into the magnetic induction !

\subsubsection{Normal multipole expansions for Cartesian components}

Assuming real values for the coefficients, $ \cplx{E}_k = E_k^n$, and taking real and imaginary parts of the resulting
equation (\ref{CEAll1}) we get the magnetic induction of normal multipoles:
\begin{eqnarray}
\label{BxyCen}
B_y^{Cen} (\eta,\psi) &=& \fr{E_1^n}{2} + \sum_{k=2}^M E_k^n  \fr{\cosh ((k-1) \eta) \ \cos ((k-1) \psi)}{\cosh((k-1) \eta_0)}  , \no \\
&&\\
B_x^{Cen} (\eta,\psi) &=& \qquad \,\  \sum_{k=2}^M E_k^n  \fr{\sinh ((k-1) \eta) \ \sin (k-1) \psi)}{\cosh((k-1) \eta_0)} \no . \no
 \end{eqnarray}
The corresponding real auxiliary function may be obtained from these formulas by
\begin{equation}
\label{BnfrXin}
\left( B_x^{Cen} (\eta,\psi), B_y^{Cen} (\eta,\psi) \right)  = - \left( \fr{\pt}{\pt \eta},  \fr{\pt}{\pt \psi} \right) \Xi^{Cen} (\eta,\psi)
\end{equation}
 or from the imaginary part of $\bfXi^{e} (\bfw)$,
[Eq.~\eqref{CEXAll}]:
\begin{eqnarray}
\label{XiElln}
&&\Xi^{Cen} (\eta,\psi) = - \ \fr{E_1^n}{2} \ \psi \  -\\
\nonumber  
&&
\ \ -  \sum_{k=2}^M \fr{1}{k-1}  E_k^n \fr{\cosh ((k-1) \eta)  \sin ((k-1) \psi)}{\cosh((k-1) \eta_0)} .
\end{eqnarray}

\subsubsection{Skew multipole expansions for Cartesian components}

Assuming imaginary values for the coefficients, $ \bfE_k = \imath E_k^s$, and taking real and imaginary parts of the resulting
equation~\eqref{CEAll1} we get the magnetic induction of skew multipoles:
\begin{eqnarray}
\label{BxyCes}
B_y^{Ces} (\eta,\psi) &=& \quad \ - \  \sum_{k=2}^M E_k^s  \fr{\sinh ((k-1) \eta) \ \sin ((k-1) \psi)}{\cosh((k-1) \eta_0)} ,\no \\
&&\\
B_x^{Ces} (\eta,\psi) &=& \fr{E_1^s}{2} + \sum_{k=2}^M E_k^s  \fr{\cosh ((k-1) \eta) \ \cos ((k-1) \psi)}{\cosh((k-1) \eta_0)}  .\no
 \end{eqnarray}
The corresponding real auxiliary function may be obtained from these formulas by
\begin{equation}
\label{BsfrXis}
\left( B_x^{Ces} \eta,\psi), B_y^{Ces} (\eta,\psi) \right) = - \left( \fr{\pt}{\pt \eta},  \fr{\pt}{\pt \psi} \right) \Xi^{Ces} (\eta,\psi)
\end{equation}
 or from the real part of $\bfXi^{Ce} (\bfw)$ (after $ \bfE_k = \imath E_k^s$ has been inserted !),
[Eq.~\eqref{CEXAll}]:
\begin{eqnarray}
\label{XiElls}
\Xi^{Ces} (\eta,\psi) &=& - \ \fr{E_1^s}{2} \ \eta \ \\\nonumber
&&- \  \sum_{k=2}^M \fr{1}{k-1} \ E_k^s \fr{\sinh ((k-1) \eta) \ \cos ((k-1) \psi)}{\cosh((k-1) \eta_0)} .
\end{eqnarray}

\subsubsection{Elliptic multipole field expansions for elliptic components}

The field expansions above were for Cartesian components which depend on elliptic coordinates.  
These are the quantities used in the fits and the evaluations. For applications, the elliptic components are now 
derived according to the rules of vector analysis in curvilinear coordinates. 
For that the Cartesian components may be transformed to components $B_\eta, B_\psi$ by Eq.~\eqref{Tftaxae}. 
The corresponding potential may be found with the help of 
\begin{equation}
\bfB^e =  \ - \mbox{grad} \Phi^{e}, \quad  (B_\eta, B_\psi) = \ - \ \fr{1}{h_t} \ \left( \fr{\pt \Phi^e}{\pt \eta}, \fr{\pt \Phi^e}{\pt \psi} \right).
\end{equation}

\paragraph{Normal multipoles}
We define new expansion coefficients
\begin{equation}
\label{ECn}
\bE_k^n \ = \ E_k^n / \cosh(k \eta_0)
\end{equation}
in Eqs.~\eqref{BxyCen} to get simpler expressions. With these and with the transformations Eqs.~\eqref{Tftaxae} we get:
\begin{widetext}
  \begin{eqnarray}
\label{bene}
B^{en}_\eta (\eta, \psi) &=& \fr{1}{2 h_t} \bigg[ \bE_1^n \cosh\eta \ \sin\psi \ + \ \bE_2^n \cosh (2 \eta ) \ \sin (2 \psi ) \  + \  \\
&&  + \ \sum_{k=2}^{M-1} \bE_{k+1}^n  \Big[ \cosh[(k+1)\eta] \sin[(k+1)\psi] - \cosh [(k-1)\eta] \sin [(k-1)\psi ]  \Big] \bigg] \nonumber\\
&=& - \frac{1}{h_t} \frac{\pt \Phi^{en}}{\pt \eta} \no\\
\label{benp}
B^{en}_\psi (\eta, \psi) &=& \fr{1}{2 h_t} \bigg[ \bE_1^n \sinh\eta \ \cos\psi \ + \ \bE_2^n \sinh (2 \eta ) \ \cos (2 \psi ) \  + \  \\
&&  + \ \sum_{k=2}^{M-1} \bE_k+1^n  \Big[ \sinh [(k+1)\eta ] \cos [(k+1)\psi ] - \sinh [(k-1)\eta] \cos[(k-1)\psi] \Big] \bigg] \nonumber\\
&=& - \frac{1}{h_t} \frac{\pt \Phi^{en}}{\pt \psi} . \no
\end{eqnarray}
\begin{eqnarray}
\label{eePhin}
\Phi^{en} (\eta, \psi) &=& -  \fr{1}{2} \bigg[ \bE_1^n \sinh\eta \ \sin\psi \ + \ \bE_2^n \ \frac{1}{2} \sinh (2 \eta ) \ \sin (2 \psi ) \  + \ \\
&& \qquad + \ \sum_{k=2}^{M-1} \bE_{k+1}^n  \Big[ \frac{ \sinh[(k+1)\eta] \sin[(k+1)\psi]}{k + 1}  - \frac{ \sinh [(k-1)\eta] \sin [(k-1)\psi ]}{k - 1}  \Big] \bigg] \no
\end{eqnarray}

\paragraph{Skew multipoles}
From Eq.~\eqref{BxyCes}, which were obtained with $\bfE_k = \imath E_k^s$, with 
\begin{equation}
\label{ECs}
\bE_k^s \ = \ E_k^s / \cosh(k \eta_0)
\end{equation}
and with Eq.~\eqref{Tftaxae} we get:
  \begin{eqnarray}
\label{bese}
B^{es}_\eta (\eta, \psi) &=& \fr{1}{2 h_t} \bigg[ \bE_1^s \sinh\eta \ \cos\psi \ + \ \bE_2^s \sinh (2 \eta ) \ \cos (2 \psi ) \  + \  \\
&&  \qquad + \ \sum_{k=2}^{M-1} \bE_{k+1}^s  \Big[ \sinh [(k+1)\eta ] \cos [(k+1)\psi ] - \sinh [(k-1)\eta ] \cos [(k-1)\psi ] \Big] \bigg] \nonumber\\
&=& - \frac{1}{h_t} \frac{\pt \Phi^{es}}{\pt \eta} , \no\\
\label{besp}
B^{es}_\psi (\eta, \psi) &=&- \ \fr{1}{2 h_t} \bigg[ \bE_1^s \cosh\eta \ \sin\psi \ + \ \bE_2^s \cosh (2 \eta ) \ \sin (2 \psi ) \  + \ \\
&&  \qquad + \ \sum_{k=2}^{M-1} \bE_{k+1}^s  \Big[ \cosh [(k+1)\eta ] \sin [(k+1)\psi ] - \cosh [(k-1)\eta ] \sin [(k-1)\psi ] \Big] \bigg] \nonumber \\
&=& - \frac{1}{h_t} \frac{\pt \Phi^{es}}{\pt \psi}. \no
\end{eqnarray}
\begin{eqnarray}
\Phi^{es} (\eta, \psi) &=& - \fr{1}{2} \bigg[ \bE_1^s \cosh\eta \ \cos\psi \ + \ \bE_2^s \ \frac{1}{2} \cosh (2 \eta ) \ \cos (2 \psi ) \  + \ \\
&&  \qquad + \ \sum_{k=2}^{M-1} \bE_{k+1}^s  \Big[ \frac{ \cosh [(k+1)\eta ] \cos [(k+1)\psi ]}{k +1} - \frac{ \cosh [(k-1)\eta ] \cos [(k-1)\psi ]}{k - 1} \Big] \bigg] . \no
\label{eePhis}
\end{eqnarray}
\end{widetext}

\subsubsection{Complex potential for normal and skew elliptic multipoles}
These series may be combined to one complex series for $B_\bfw^e = B^{e}_\psi + \imath B^{e}_\eta$.  The resulting series 
may be integrated to give a series for the complex potential:
\begin{eqnarray}
\Phi^{e} (\bfw) &= - \  \fr{\imath}{2} \bigg[&  \bfE_1 \ \cosh \cplx{w} \ + \ \fr{1}{2} \bfE_2 \ \cosh (2 \cplx{w}) \ + \\ \no  
&& \qquad +  \sum_{k=2}^{M-1}  \bfE_{k+1} \big(\fr{1}{k+1} \cosh [(k + 1) \cplx{w}]\\\no&&\quad\quad -  \fr{1}{k-1} \cosh [(k - 1) \cplx{w}] \big) \bigg] . 
\end{eqnarray}

%
%


\subsection{Toroidal Multipoles}

Toroidal multipole are useful for representing the potential and field in the gap of curved magnets.
Up to now, plane circular multipoles have been used to represent these quantities in most cases.
But the toroidal multipoles are better adapted to the geometry. Toroidal coordinates (see e.g.\cite{MS}, Fig.4.4)
have an infinite set of nested tori as one set of coordinate surfaces. In these coordinates the potential
can be solved exactly by R-separation. An example of this has been given by \cite{auchmann:compumag-field-harmonics},
\cite{global-toroidal-brouwer-pac13}.

Local toroidal coordinates have also an infinite set of nested tori as one set of coordinate surfaces. In these
coordinates the  potential equation can be solved by approximate R-separation for slender rings. But these
solutions are nearer to the plane circular multipoles than the particular solutions of the potential equation
in the standard torus coordinates.  Since the condition of "slenderness" is met in most practical cases we
prefer to use the local toroidal coordinates. In addition this approach can be generalised to slender 
rings with elliptical cross section. 

The toroidal circular multipoles are a complete set of solutions as are the plane circular multipoles. 
So each of these systems may be employed to expand the potential and the field in the gap of a magnet. 
However, In a curved magnet  the toroidal multipoles give a more accurate solution of the potential equation
than the plane circular multipoles.

\subsubsection{Circular toroidal multipoles}
\label{sec:toroidal-circular-multipoles}
\begin{figure}[b]
\centering\includegraphics[width=8cm]{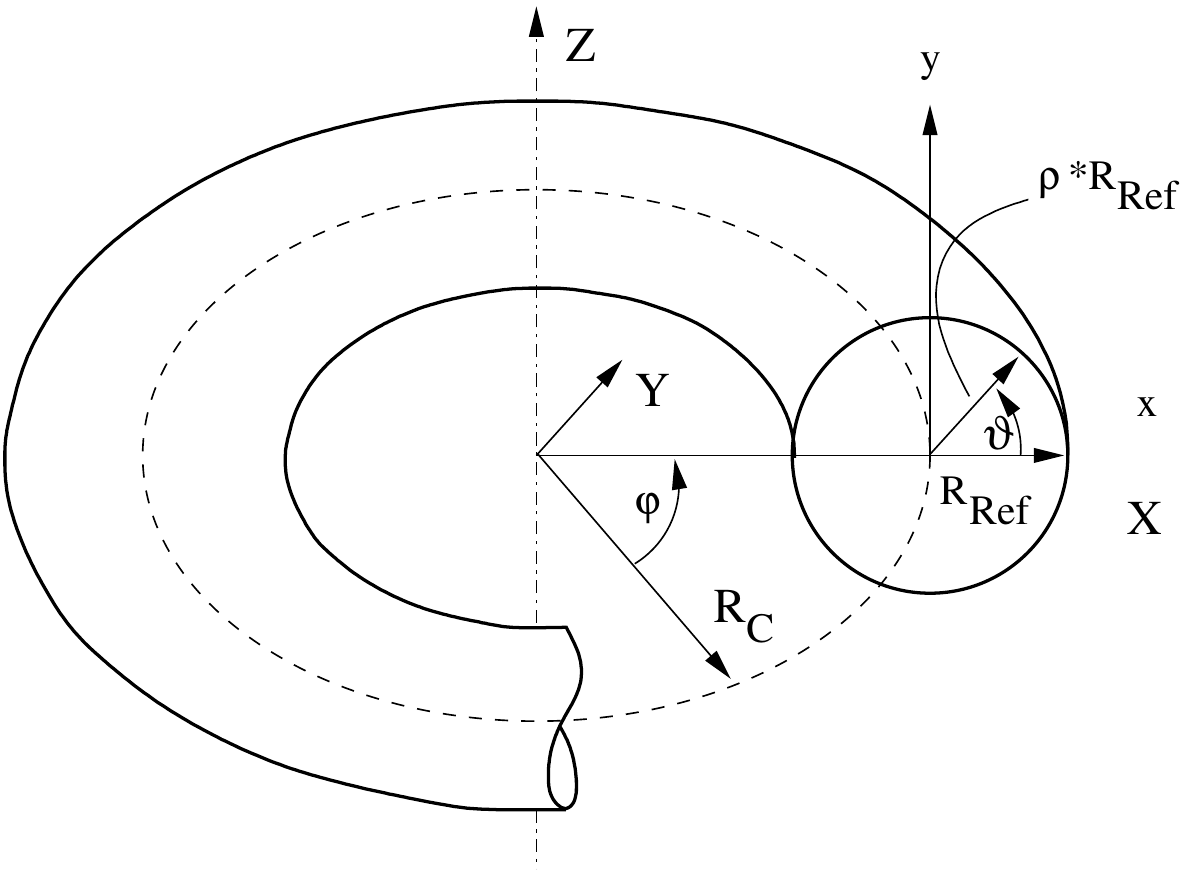}
\caption{Relative local toroidal coordinates $\rho, \ \vth , \ \vph$. 
  Graphic courtesy of B. Seiwald \protect\cite{Seiwald:phd}.}
\label{figloctorco}
\end{figure}

\paragraph{Circular local toroidal coordinates}
\label{sec:local-toroidal-coordinates} 
These coordinates start from a system of plane polar coordinates  $r, \theta$ with $x = r \cos\theta, \ y = r \sin\theta.$
 At first this system is standing in a three-dimensional space $X, Y, Z$ such that the $x-$ and the $X-$ and the
 $y-$ and the $Z-$axis are the same. Then the polar system is shifted by $R_c$ along the $X$-axis. Finally, the
 shifted system is rotated by an angle $\phi$ around the original $Z-$axis. The circles of the shifted polar coordinate
 system give tori, all having the same centre circle $\sqrt{X^2 + Y^2} = R_c, \ Z = 0$ 
(see also Fig.~\ref{figloctorco}, \cite{LCK,PlasmaBook,pbep:compel}).
 The smaller radius of the torus segment, $\rref$, must be smaller than the curvature radius $R_c$. 
 The coordinates can only be used in the full interior of the reference torus :  $\sqrt{x^2 + y^2} \leq \rref < R_c$.

It is convenient to use a dimensionless quasi-radius $\rho = r/\rref$. Then the local 
torus coordinates $\rref \rho, \theta, \phi $ are defined as:
\begin{eqnarray}
X &=& R_c \ h \ \cos\phi, \\
Y &=& R_c \ h  \ \sin\phi, \\
Z &=& \rref \sin\theta ; \\
h  &=& 1 + \eps \ \rho \cos\theta.
\end{eqnarray}
\begin{equation}
0 \leq \eps := \rref/R_c < 1
\end{equation}
is called the inverse aspect ratio. It is smaller than unity; in most accelerators even much smaller.
$0 \leq \rho \leq 1$ is denoted as the dimensionless quasi-radius; $ - \pi \leq \theta \leq \pi$ as the poloidal angle;
$- \phi_0 \leq \phi \leq \phi_0$ as the toroidal (= azimuthal) angle.

\paragraph{Circular local toroidal multipoles}
 
Here only potentials, so fields, are considered, which are toroidally uniform;
thus these quantities are the same in every cross section $\phi =$ const.;
they do not depend on the toroidal coordinate $\phi$, depend only on the dimensionless quasi-radius $\rho$ 
and on the poloidal coordinate $\theta$.  The corresponding potential is a solution of the following potential
equation (some unessential constant factors have been omitted):
\begin{eqnarray}
\label{PotEqLoCir}
&&\Delta\Psi \!=\!\\
\nonumber
 &&\left[ \fr{\pt^2}{\pt\rho^2} + \fr{1}{\rho} \fr{\pt}{\pt \rho} +  \fr{1}{\rho^2}  \fr{\pt^2}{\pt\theta^2}  +  
\fr{\eps}{h} \left( \cos\theta \fr{\pt}{\pt \rho} - \sin\theta \fr{1}{\rho} \fr{\pt}{\pt \theta}  \right)\right] \Psi = 0 \\
\label{PotEqLoCirSep}
 && \fr{1}{\sqrt{h}}  \left[ \fr{\pt^2}{\pt\rho^2} + \fr{1}{\rho} \fr{\pt}{\pt \rho} +  \fr{1}{\rho^2} \fr{\pt^2}{\pt\theta^2} + 
 \fr{\eps^2}{h^2} \right](\sqrt{h} \Psi) = 0. 
\end{eqnarray}
In going to the last equation the dependent variable $\Psi$ has been replaced with $\sqrt{h} \Psi$; 
this substitution entails the change of the perturbation from a power series in $\eps$ starting with a 
 term linear in $\eps$ into a new one starting with a quadratic term:
\begin{equation}
\fr{\eps^2}{h^2} = \eps^2 \ (1 + \eps \ \rho \cos\theta)^{-2} = \eps^2 (1 - 2 \eps \ \rho \ \cos\theta + ... ).
\end{equation}
These two interrelated changes are the essence of the R-separation. In the local torus coordinates the term
$1/h$ or $1/h^2$ prevents the separation of variables. But dropping $\eps^2/h^2$ or approximating 
$h^2 \approx 1$ in Eq.~\eqref{PotEqLoCirSep}
both give separable equations. We call this \textbf{Approximate R-separation.}

When  $\eps^2/h^2$ is dropped in Eq.~\eqref{PotEqLoCirSep} the resulting differential operator is that of the potential equation 
in polar coordinates $\rho, \theta$. With a solution of the latter, $\Phi (\rho, \theta),$
we get an approximate solution of the former which is accurate to the 
first order in $\eps$:
\begin{eqnarray}
\label{EllSoPsi}
\Psi (\rho, \theta) \ &=& \ (1 + \eps \rho \cos\theta)^{-1/2} \ \Phi (\rho, \theta) \ + \ O(\eps^2) \\\nonumber
&=& \ (1 - \fr{1}{2}  \eps \rho \cos\theta) \ \Phi (\rho, \theta) \ + \ O(\eps^2) . 
\end{eqnarray}
Identifying $\Phi (\rho, \theta)$ with $\Phi_m =  \rho^m \ e^{\imath m \theta}, \ m = 0, 1, 2 ,...$ we get:
\begin{eqnarray}
\Psi_m &=&  h^{-1/2} \ \rho^m \ e^{i m \theta} \ \\\nonumber 
&\approx& \  \rho^m \ e^{\imath m \theta} \big(1 - \fr{ \eps}{4} \  \rho (e^{\imath \theta} +  e^{-\imath \theta})\big) \\\nonumber
&=& \rho^m \ e^{\imath m \theta} -  \fr{ \eps}{4} \rho^{m + 1} \big( e^{\imath (m + 1) \theta} + e^{\imath (m - 1) \theta} \big).
\end{eqnarray}
In the first order approximation the curvature term
adds two terms: $\rho^{m + 1} \ e^{\imath (m + 1) \theta} $ is again a solution of the Laplacian in polar coordinates.
At the contrary, the last term alone, i.e. $\rho^{m + 1} e^{\imath (m - 1) \theta}$, is not a solution of that operator.
Inserting $\Psi_m$, Eq.~\eqref{EllSoPsi}, into Eq.~\eqref{PotEqLoCir} gives $\Delta \Psi_m = 0 + O(\eps^2)$.
So it is verified that  $\Psi_m$ is an approximate solution of the potential equation in local toroidal coordinates.
 
Here it is convenient to define local Cartesian coordinates $x, y$ in the cross section $\phi =$ const.:
\begin{equation}
x = \rref \ \rho \ \cos\theta, \qquad y =  \rref \ \rho \ \sin\theta; \quad \bfz = x + \imath y .
\end{equation}
and to transform  $\Psi_m$ accordingly:
\begin{eqnarray}
\label{Tor:Ph1xy} 
\Psi_m &= 
- \frac{1}{m} \ \Bigg\{&
\underbrace{\left(\frac{\cz}{\rr} \right)^{|m|}}_{T_0}   -  
\\ \nonumber
&&
- \frac{\eps}{4} 
\Bigg[
  \underbrace{\left( \frac{\cz}{\rr} \right)^{|m|+1}}_{T_1} + 
  \underbrace{\frac{\left|\cz\right|^2}{\rr^2} \left(\frac{ \cz}{\rr}\ \right)^{|m|-1}}_{T_2} 
\Bigg] \Bigg\}.  
\end{eqnarray}
The presence of $|\bfz|^2$ indicates that $\Psi_m$ is not a complex analytic function of $\bfz$.
These local Cartesian coordinates correspond to the Cartesian coordinates used for 
Plane Circular Multipoles in section~\ref{sec:circular-multipoles}.  Multiplying $ \Phi^{C} (x, y)$ by 
\begin{eqnarray}
\nonumber
h^{-1/2} &=& (1 +  \eps \ x/\rref)^{-1/2} \approx 1 - \frac{1}{2} \ \eps \ \frac{x}{\rref} \ + \ O(\eps^2); \\
\nonumber
S &:=& 1 \ - \  \frac{1}{2} \ \eps \ \frac{x}{\rref}
\end{eqnarray}
gives the wanted approximate solution $\Psi (x, y)$ of the potential equation in local toroidal coordinates.
In particular, by this multiplication  $ \Phi^{Cn}$ [Eq.~\eqref{PhCn}], or $\Phi^{Cs}$ [Eq.~\eqref{PhCs}], will be transformed to 
normal or skew potentials  $ \Psi^{Cn} (\rho, \theta)$ or $ \Psi^{Cs} (\rho, \theta)$ for curved magnets:
\begin{eqnarray}
\label{TrPhiPsi}
\Psi^{C\alpha}_m (x, y)  &=&  S  \Phi^{C\alpha}_m (x, y)\\ 
\nonumber
& =& \Phi^{C\alpha}_m (x, y)  -  \frac{1}{2}  \eps  \frac{x}{\rref} \Phi^{C\alpha}_m (x, y) ;
\quad \alpha = n, s
\end{eqnarray}
are the normal and the skew toroidal multipoles accurate to the first order in $\eps$.

The potentials  $ \Phi^{Cn}$ [Eq.~\eqref{PhCn}] or $\Phi^{Cs}$ [Eq.~\eqref{PhCs}] could contain a constant term $C_0$,
which is not included in the aforementioned equations. Such a constant does not give a contribution to the magnetic 
induction in the cylindrical case. However, in Eq.~\eqref{TrPhiPsi} it would lead to a non-vanishing toroidal function
\[
\Psi^{C\alpha}_0 (x, y) \ = \ S \ C_0 \ = \ C_0  - \frac{1}{2} \ \eps \ \frac{x}{\rref} \ C_0 .
\]
There is a good reason for dropping such a basis function:
There is a one-to-one correspondence between the cylindrical and the toroidal basis functions,
so between the elements of the two sets of basis functions. The former set is complete, so is the new one.
The reference volume of the first set is a straight finite cylinder. The second reference volume is a segment
of a torus, which may be obtained from the first one by simple bending it. This operation does not change
the topology of the volume. Only a real change of the topological connectivities would provoke the need 
for additional basis functions. 

Vectorial basis functions are needed for expanding the magnetic field. These are obtained by taking the gradients
of the potentials
\begin{eqnarray}
\label{Tor:VFxy}
&&\vec{T}_m^{C\alpha} (x, y)  =  - \grad \Psi^{C\alpha}_m (x, y) \\ \nonumber
&& = 
\ - \grad \Phi^{C\alpha}_m (x, y) \ +  \frac{1}{2} \ \eps \ \grad\left(  \frac{x}{\rref} \ \Phi^{C\alpha}_m (x, y) \right), 
\alpha = n, s ; \\
\nonumber
&&= 
\left(
\begin{array}{@{\hspace{.1em}}c@{\hspace{.1em}}c@{\hspace{.1em}}c@{\hspace{0em}}}
 - \frac{\pt \Phi^{C\alpha}_m (x, y)}{\pt x} &+  \frac{1}{2}  \eps   \frac{x}{\rref}  \frac{\pt \Phi^{C\alpha}_m (x, y)}{\pt x} &
  +  \frac{1}{2}  \eps   \frac{1}{\rref}  \Phi^{C\alpha}_m (x, y)  \\[1mm]
 - \frac{\pt \Phi^{C\alpha}_m (x, y)}{\pt y} &+   \frac{1}{2}  \eps  \frac{x}{\rref}  \frac{\pt \Phi^{C\alpha}_m (x, y)}{\pt y} &
\end{array}
\right),
\end{eqnarray}
substituting $\Phi^{Cn}$ [Eq.~\eqref{PhCn}] or $\Phi^{Cs}$ [Eq.~\eqref{PhCs}] for  $\Phi^{C\alpha}$.
The  vector basis functions $\vbT_m^{(n)}$ for the normal components $B_m$ are then obtained using $\Phi^{Cn}$ and 
vector basis functions $\vbT_m^{(s)}$ for the skew components $A_m$ using $\Phi^{Cs}$. 
Any magnetic induction may then expanded in the following way:
\begin{equation}
\label{Tor:BExp}
\vec{B}  (x, y)  = \sum_{m=1}^M \left[B_m \  \vbT_m^{(n)}(x, y) \ +  A_m \ \vbT_m^{(s)}(x, y)\right]\,.
\end{equation} 
A complex presentation of the field allows obtaining a closed expression for the magnetic field introducing 
basis functions $\cplx{\vbT^{(n)}_m}$, $\cplx{\vbT^{(s)}_m}$ which fulfil
\begin{equation}
\label{Tor:BExpComplex}
  \cplx{B}  (\cplx{z})  = \sum_{m=1}^M \left[B_m \ \cplx{\vbT^{(n)}_m}(\cplx{z})  \ +  \imath A_m \ \cplx{\vbT^{(s)}_m}(\cplx{z})\right]\,.
\end{equation} 
Comparing Eq.~\eqref{Tor:BExp} to Eq.~\eqref{Tor:BExpComplex} one can see that 
 $\cplx{\vbT^{(n)}_m}$ and $\cplx{\vbT^{(n)}_s}$ are related to ${\vbT^{(n)}_m}$, ${\vbT^{(s)}_m}$ by
\begin{eqnarray}
\label{Tor:VFnxyC}
  \vbT_m^{(n)} (x, y) &:=&\re{\left(\      \cplx{\vbT^{(n)}_m}(\cplx{z})\right)}\uv{y} + \im{\left(\      \cplx{\vbT^{(n)}_m}(\cplx{z})\right)}\uv{x},    \\  
\label{Tor:VFsxyC}  
  \vbT_m^{(s)} (x, y) &:=&\re{\left(\imath \cplx{\vbT^{(s)}_m}(\cplx{z})\right)}\uv{y} + \im{\left(\imath \cplx{\vbT^{(s)}_m}(\cplx{z})\right)}\uv{x}. 
\end{eqnarray}
or expressing the complex functions by $\cplx{\vbT^{(n)}_m}$ and $\cplx{\vbT^{(n)}_s}$
\newcommand{\tzarg}{\left(\re\left(\cplx{z}\right),\im\left(\cplx{z}\right)\right)}%
\begin{eqnarray}
\label{Tor:VFnxy}
  \cplx{\vbT_m^{(n)}} (\cplx{z}) & = &  \left(\vbT_m^{(n)}\tzarg\cdot\uv{y}\right) \\\nonumber&& + \left(\vbT^{(n)}_m\tzarg\cdot\uv{x}\right) \imath \\  
\label{Tor:VFsxy}
  \cplx{\vbT_m^{(s)}} (\cplx{z}) & = &  \left(\vbT_m^{(s)}\tzarg\cdot\uv{y}\right) \\\nonumber&& + \left(\vbT^{(s)}_m\tzarg\cdot\uv{x}\right) \imath
\end{eqnarray}    

Cylindric circular multipoles are only satisfying the potential
equations if $m > 0$ (see e.g. \cite{jain98:_magnet}).
The same requirement is now imposed on $m$ here as 
otherwise the aforementioned condition will not be fulfilled
for $\eps = 0$. Therefore it is assumed that $m > 0$
holds also for the toroidal circular multipoles.
Two methods can now be used to derive these complex functions, both with their merits. The first method
uses the fact that the cylindric circular multipoles and their potential can be expressed by the forms
given in Eq.~\eqref{BCn} and Eq.~\eqref{PhCn} for the normal multipoles and by Eq.~\eqref{BCs} and Eq.~\eqref{PhCs} for 
the skew ones. Inserting these expressions in Eq.~\eqref{Tor:VFxy} one obtains for the normal components
\begin{widetext}
  \begin{equation}   
 \vbT_m^{(n)} (x, y) =     \left(
\begin{array}{ccc}
   \im\left(\ \cplx{B^{C}}\right) &-  \frac{1}{2}  \eps   \frac{x}{\rref}     \im\left( \cplx{B^{C}}\right)&
    -  \frac{1}{2}  \eps   \frac{1}{\rref}    \im\left( \cplx{\Phi^{Cn}}\right)  \\[1mm]
    \re\left( \cplx{B^{C}}\right) &-  \frac{1}{2}  \eps   \frac{x}{\rref}     \re\left( \cplx{B^{C}}\right)&
 \\[1mm]
\end{array}  
\right)
=
\left(
\begin{array}{c}
\im\left[\cplx{\ \vbT^{(n)}_m}\right]\\
\re\left[\cplx{\ \vbT^{(n)}_m}\right]
\end{array}  
\right)
  \end{equation}
and for the skew components
  \begin{equation}
 \vbT_m^{(s)} (x, y) = \left(
\begin{array}{ccc}
    \im\left(\cplx{\imath B^{C}}\right) &-  \frac{1}{2} \ \eps \  \frac{x}{\rref} \    \im\left(\imath \cplx{B^{C}}\right)&
 \ - \ \frac{1}{2} \ \eps \  \frac{1}{\rref}     \im\left(\imath \cplx{\Phi^{Cn}}\right)  \\[1mm]
    \re\left(\cplx{\imath B^{C}}\right) &-  \frac{1}{2} \ \eps \  \frac{x}{\rref} \    \re\left(\imath \cplx{B^{C}}\right)&
 \\
\end{array}      
\right)
=
\left(
\begin{array}{c}
\im\left[\cplx{\imath \vbT^{(s)}_m}\right]\\
\re\left[\cplx{\imath \vbT^{(s)}_m}\right]
\end{array}  
\right)
.
\end{equation}
These can  now be combined to  complex ones.
Using 
\begin{equation}
   \im\left(\ \cplx{\Phi^{Cn}}\right) = \frac 1 n \im\left(\frac{\cplx{z}}{\rref} \cplx{B^{C}} \right)
   \qquad
   \im\left(\imath \cplx{\Phi^{Cn}}\right) = \frac 1 n \im\left(\imath \frac{\cplx{z}}{\rref} \cplx{B^{C}} \right)
\end{equation}
one obtains
\begin{equation}
\label{eq:local-toroidal-multipoles-basis-functions-solution1}
  \left(
    \begin{array}{c}
      \cplx{\vbT_m^{(n)}} (\cplx{z})\\
      \cplx{\vbT_m^{(s)}} (\cplx{z})
    \end{array}  
  \right)
  = 
  \left(\frac{\cplx{z}}{\rref}\right)^{n-1} \left(1 - \eps \frac 1 2 \frac{\re{\left(\cplx{z}\right)}}{\rref}\right) - 
  \eps \frac{2}{n}  
  \left(
    \begin{array}{l}
      \im\left[\left(\frac{\cplx{z}}{\rref}\right)^{n}\right]  \imath\\\
      \re\left[\left(\frac{\cplx{z}}{\rref}\right)^{n}\right]
    \end{array}  
    \right)
  ,
\end{equation}
or  substiuting $\cplx{z}/\rref$ with $\rho e^{\imath \vartheta}$ one gets
\begin{eqnarray}
  \left(
    \begin{array}{c}
      \cplx{\vbT_m^{(n)}} (\cplx{z})\\
      \cplx{\vbT_m^{(s)}} (\cplx{z})
    \end{array}  
  \right)
  &=& \rho^{n-1}e^{\imath (n - 1) \vartheta} \left(1 - \eps\rho \cos(\vartheta)\right) - 
  \eps \frac{\rho^n}{2 n} 
  \left(
    \begin{array}{l}
      \sin\left(n \vartheta\right)  \imath\\\
      \cos\left(n \vartheta\right)\\
    \end{array}  
  \right)  \\
  &=& 
  \rho^{n-1}  \left(
    e^{\imath  (n - 1) \vartheta} -     \frac{\eps\rho}{2}\left[      e^{\imath  (n - 1) \vartheta} 
      \cos(\vartheta) +  \frac 1 n 
  \left(
    \begin{array}{l}
      \sin\left(n \vartheta\right)  \imath\\
      \cos\left(n \vartheta\right)\\
    \end{array}  
  \right)  
\right]
\right)
\end{eqnarray}
in polar representation.  
\end{widetext}

The second approach is using the expression Eq.~\eqref{Tor:Ph1xy} and derive the potential 
for the normal and skew multipoles as given in Eq.~\eqref{BCn} and Eq.~\eqref{BCs}, as 
the first two term $T_0$  and $T_1$ are analytic and a complex representation of the 
solution facilitates interpretation.
Given that $\abs{\cplx{z}}$ is part of the ansatz, the calculation can 
not be performed using complex coordinates safely. Instead these 
calculations were performed by two steps:
\begin{itemize}
\item At first all coordinates were substituted by their real
  values. Then the calculations were performed given by
  (\ref{Tor:VFxy}).
\item Based on the results complex representations were deduced and
  Eqs.~\eqref{Tor:VFnxyC} and \eqref{Tor:VFsxyC} were
  applied to verify the results. These equations are chosen such that the  unperturbed term
  will produce the basis functions of conventional cylindric circular multipoles.
\end{itemize}
These terms $\vbT_m$ were calculated to high order using a computer
algebra system \cite{sympy}. Based on these results a formula was obtained and checked
using \cite{sympy} and Mathematica\texttrademark.
Term 2 of Eq.~\eqref{Tor:Ph1xy} yields the following equation  
\begin{widetext}  
{
 \begin{equation}
     \left(
   \begin{array}[c]{@{\hspace{0em}}c@{\hspace{0em}}}
   \cplx{\vbT^{(n)}_m}\\
   \cplx{\vbT^{(s)}_m}
 \end{array}
       \right)
      = \zs^2 \left(1 - m\right) 
      \zs^{m - 2}
      +  \frac{2}{\rref^{m-1}} 
      \left(
     \begin{array}[c]{c}
        y \re{\left(\cplx{z}^{m-1} \imath\right)}  - \imath x \im{\left(\cplx{z}^{m -1}\right)}\\
      -y \im{\left(\cplx{z}^{m-1} \imath\right)} - \imath x \re{\left(\cplx{z}^{m -1}\right)}\\       
     \end{array}
     \right),
 \end{equation}
}
which can be reformulated to 
{
\small
 \begin{equation}
     \label{eq:circular-toroidal-basis-function-short-form}
     \left(
   \begin{array}[c]{@{\hspace{0em}}c@{\hspace{0em}}}
   \cplx{\vbT^{(n)}_m}\\
   \cplx{\vbT^{(s)}_m}
 \end{array}
       \right)
      = 
      \left(\frac{\abs{\cplx{z}}}{\rref}\right)^2 
      \frac{\left(m - 1\right)}{m}
      \zs^{m - 2}
      +  \frac{2 \conj{z}}{m \rref} 
      \left(
     \begin{array}[c]{@{\hspace{0em}}c@{\hspace{.1em}}c@{\hspace{0em}}}
          \imath & \im{\left(\left({\cplx{z}}/{\rref}\right)^{m-1} \right)} \\
                      & \re{\left(\left({\cplx{z}}/{\rref}\right)^{m-1}\right)} \\       
     \end{array}
     \right),
 \end{equation}
}
with $\conj{z} = x + \imath y$.
The first basis functions for $T_2$ are given in Table~\ref{tab:circular-toroidal-basis-terms}.
 \todo{ are these correct?} 
\begin{table}
  \caption{Potential and basis functions for term $T_2$.}
  \label{tab:circular-toroidal-basis-terms}
  \centering  
  \begin{ruledtabular}
      \begin{tabular}[c]{%
      l
      c
      c
      c
    }
    $m$ &
               \multicolumn{1}{c}{$\Phi$} &
               \multicolumn{1}{c}{$B_x$} &
               \multicolumn{1}{c}{$B_y$} \\
\hline
\multicolumn{4}{l}{normal}\\

     1 & ${0}$ 
& $0$ & $0$\\

     2 & ${- \frac{x^{2} y}{R_{{Ref}}^{3}} - \frac{y^{3}}{R_{{Ref}}^{3}}}$ 
& $\frac{x y}{R_{{Ref}}^{2}}$ & $\frac{x^{2} + 3 y^{2}}{2 R_{{Ref}}^{2}}$\\

     3 & ${- 2 \frac{x^{3} y}{R_{{Ref}}^{4}} - 2 \frac{x y^{3}}{R_{{Ref}}^{4}}}$ 
& $\frac{2}{3} \frac{y \left(3 x^{2} + y^{2}\right)}{R_{{Ref}}^{3}}$ & $\frac{2}{3} \frac{x \left(x^{2} + 3 y^{2}\right)}{R_{{Ref}}^{3}}$\\

     4 & ${- 3 \frac{x^{4} y}{R_{{Ref}}^{5}} - 2 \frac{x^{2} y^{3}}{R_{{Ref}}^{5}} + \frac{y^{5}}{R_{{Ref}}^{5}}}$ 
& $\frac{x y \left(3 x^{2} + y^{2}\right)}{R_{{Ref}}^{4}}$ & $\frac{3 x^{4} + 6 x^{2} y^{2} - 5 y^{4}}{4 R_{{Ref}}^{4}}$\\

     5 & ${- 4 \frac{x^{5} y}{R_{{Ref}}^{6}} + 4 \frac{x y^{5}}{R_{{Ref}}^{6}}}$ 
& $\frac{4}{5} \frac{y \left(5 x^{4} - y^{4}\right)}{R_{{Ref}}^{5}}$ & $\frac{4}{5} \frac{x \left(x^{4} - 5 y^{4}\right)}{R_{{Ref}}^{5}}$\\
\hline
\multicolumn{4}{l}{skew}\\

     1 & ${- \frac{x^{2}}{R_{{Ref}}^{2}} - \frac{y^{2}}{R_{{Ref}}^{2}}}$ 
& $2 \frac{x}{R_{{Ref}}}$ & $2 \frac{y}{R_{{Ref}}}$\\

     2 & ${- \frac{x^{3}}{R_{{Ref}}^{3}} - \frac{x y^{2}}{R_{{Ref}}^{3}}}$ 
& $\frac{3 x^{2} + y^{2}}{2 R_{{Ref}}^{2}}$ & $\frac{x y}{R_{{Ref}}^{2}}$\\

     3 & ${- \frac{x^{4}}{R_{{Ref}}^{4}} + \frac{y^{4}}{R_{{Ref}}^{4}}}$ 
& $\frac{4}{3} \frac{x^{3}}{R_{{Ref}}^{3}}$ & $- \frac{4}{3} \frac{y^{3}}{R_{{Ref}}^{3}}$\\

     4 & ${- \frac{x^{5}}{R_{{Ref}}^{5}} + 2 \frac{x^{3} y^{2}}{R_{{Ref}}^{5}} + 3 \frac{x y^{4}}{R_{{Ref}}^{5}}}$ 
& $\frac{5 x^{4} - 6 x^{2} y^{2} - 3 y^{4}}{4 R_{{Ref}}^{4}}$ & $\frac{x y \left(- x^{2} - 3 y^{2}\right)}{R_{{Ref}}^{4}}$\\

     5 & ${- \frac{x^{6}}{R_{{Ref}}^{6}} + 5 \frac{x^{4} y^{2}}{R_{{Ref}}^{6}} + 5 \frac{x^{2} y^{4}}{R_{{Ref}}^{6}} - \frac{y^{6}}{R_{{Ref}}^{6}}}$ 
& $\frac{6 x^{5} - 20 x^{3} y^{2} - 10 x y^{4}}{5 R_{{Ref}}^{5}}$ & $\frac{2}{5} \frac{y \left(- 5 x^{4} - 10 x^{2} y^{2} + 3 y^{4}\right)}{R_{{Ref}}^{5}}$\\
  \end{tabular}
  \end{ruledtabular}
\end{table}    
For term $T_0$ and term $T_1$ one obtains
\begin{equation}
  T_0:\ \cplx{\vbT^{(n)}_m} =  \cplx{\vbT^{(s)}_m} = \zs^{m-1}
  \qquad
  T_1:\
  \cplx{\vbT^{(n)}_m} =  \cplx{\vbT^{(s)}_m} = \frac{m + 1}   m \zs^{m} .    
 \end{equation}
Thus term $T_1$ is a ``feed-up'', similar as a translation of coordinate systems gives a feed down.
The results for these terms can be combined to 
{\small
\begin{eqnarray}
\label{eq:toroidal-circular-tn-ts-cplx}
     \left(
   \begin{array}[c]{@{\hspace{0em}}c@{\hspace{0em}}}
   \cplx{\vbT^{(n)}_m}\\
   \cplx{\vbT^{(s)}_m}
 \end{array}
\right)
&=&  
\zs^{m-1}\!\!\!\!
-\\ 
\nonumber
&&
\!\!\!\! - \frac{\epsilon}{4 m}
\Bigg\{\!
  \zs^{m-2} 
  \left[
    \left(m + 1 \right) \zs^2 + 
    \left(m - 1 \right) \abs{\zs}^2
  \right]    
 + 
  \frac{2 \conj{z}}{\rref} 
      \left[
     \begin{array}[c]{@{\hspace{0em}}c@{\hspace{.1em}}c@{\hspace{0em}}}
          \imath & \im{\left(\left({\cplx{z}}/{\rref}\right)^{m-1} \right)} \\
                      & \re{\left(\left({\cplx{z}}/{\rref}\right)^{m-1}\right)} \\       
     \end{array}
     \right] 
\Bigg\} \, .
\end{eqnarray}
}
The results of the equation above were compared to 
Eq.~\eqref{eq:local-toroidal-multipoles-basis-functions-solution1} and found to be identical.
The
equation can be rewritten replacing $\left(\cplx{z}/\rref\right)$ with
$\rho e^{\imath\vartheta}$. 
{
\small
\begin{equation}
  \left(
   \begin{array}[c]{@{\hspace{0em}}c@{\hspace{0em}}}
   \cplx{\vbT^{(n)}_m}\\
   \cplx{\vbT^{(s)}_m}
 \end{array}
\right)
=
\rho^{m-1} e^{\imath (m - 1) \vartheta} 
\\
 - \frac{ \epsilon}{4 m}
 \left(
   \rho^{m+1} e^{\imath (m - 2) \vartheta}
  \left[
    \left(m + 1 \right)  e^{2 \imath \vartheta } + 
    \left(m - 1 \right) 
  \right]     
+ 2 \rho^m
  e^{- \imath \vartheta}   
      \left[
     \begin{array}[c]{@{\hspace{0em}}c@{\hspace{.1em}}c@{\hspace{0em}}}
       e^{\imath\pi/2}& \sin{\left(\left({m-1}\right) \vartheta \right)}\\
                      & \cos{\left(\left({m-1}\right) \vartheta \right)}\\       
     \end{array}
     \right]
     \right)
\, ,
\end{equation}
}
It can be further transformed to 
{
\small
\begin{equation}
  \left(
   \begin{array}[c]{@{\hspace{0em}}c@{\hspace{0em}}}
   \cplx{\vbT^{(n)}_m}\\
   \cplx{\vbT^{(s)}_m}
 \end{array}
\right)
=
\rho^{m-1} \left(e^{\imath (m - 1) \vartheta} 
\\
 - \frac{\rho \epsilon}{4 m}
 \left\{
     e^{\imath (m - 2) \vartheta}
  \left[
    \left(m + 1 \right)  e^{2 \imath \vartheta } + 
    \left(m - 1 \right) 
  \right]     
+ {2}  e^{- \imath \vartheta}   
      \left[
     \begin{array}[c]{@{\hspace{0em}}c@{\hspace{.1em}}c@{\hspace{0em}}}
       e^{\imath\pi/2}& \sin{\left(\left({m-1}\right) \vartheta \right)}\\
                      & \cos{\left(\left({m-1}\right) \vartheta \right)}\\       
     \end{array}
     \right]
     \right\}
\right)
\, ,
\end{equation}
}
and 
{
\small
\begin{equation}
  \left(
   \begin{array}[c]{@{\hspace{0em}}c@{\hspace{0em}}}
   \cplx{\vbT^{(n)}_m}\\
   \cplx{\vbT^{(s)}_m}
 \end{array}
\right)
=
\rho^{m-1} e^{\imath (m - 1)\vartheta} \left(1
\\
 - \frac{\rho \epsilon}{4 m}
 \left\{
  \left[
    \left(m + 1 \right)  e^{ \imath \vartheta } + 
    \left(m - 1 \right)  e^{-\imath  \vartheta}
  \right]     
  +  2  e^{-\imath  \vartheta}   
      \left[
     \begin{array}[c]{@{\hspace{0em}}c@{\hspace{.1em}}c@{\hspace{0em}}}
       e^{\imath\pi/2}& \sin{\left(\left({m-1}\right) \vartheta \right)}\\
                      & \cos{\left(\left({m-1}\right) \vartheta \right)}\\       
     \end{array}
     \right]
     \right\}
\right)
     \, .
\end{equation}
}
\end{widetext}

The equations above are just reformulations of
Eqs.~\eqref{eq:local-toroidal-multipoles-basis-functions-solution1} and \eqref{eq:toroidal-circular-tn-ts-cplx} 
but simplify the understanding
of the obtained results. One can see that 
\begin{itemize}
\item the whole perturbation depends linearly on the offset from the
  centre circle. 
\item The distortions decrease for higher orders of $m$.
\item The first term corresponds to a feed up, as if a
  multipole with order ${\epsilon (m +1)}/4$ was added.
  The other one corresponds to a field increasing with the distance
  from the centre circle. 

\item The last term, which is different for $\cplx{\vbT^{n}}$ and
  $\cplx{\vbT^{s}}$, is counter-rotating with the phase of the field.
Finally  $\cplx{\vbT^{s}}$ is scaled by the imaginary  component 
while $\cplx{\vbT^{s}}$ is  scaled by the real
component  of $\cplx{z}/\abs{\cplx{z}}$ in
the direction $x$.
\end{itemize}
The distortions are non harmonic solutions except for the feed up.
The last term shows that the field is decreasing with increasing $x$.
Given that the solution was obtained bending the basis functions and
thus ``the magnet'', these are not surprising. A straight air coil dipole
magnet which is bent would show similar behaviour, as the current
density decreases per length on the outside but increases at the inside.
The rotations follow similar
insight, as the field direction should change too (imagine a 
straight sextupole coil which is bent to a torus).

\subsection{Elliptic toroidal multipoles}

\subsubsection{Elliptic local toroidal coordinates}
\label{sec:local-toroidal-elliptic-coordinates}

If instead of a circle an ellipse is revolved around the major circle
(like pressing down a doughnut), an adapted coordinate system is a
system of Local Elliptic Toroidal Coordinates. Thus the minor circle
is replaced by an ellipse.

\begin{figure}[tbp]
  \centering
  \begin{tabular}[c]{@{\hspace{0em}}c@{\hspace{0em}}c@{\hspace{0em}}}
    \includegraphics[width=.5\columnwidth]{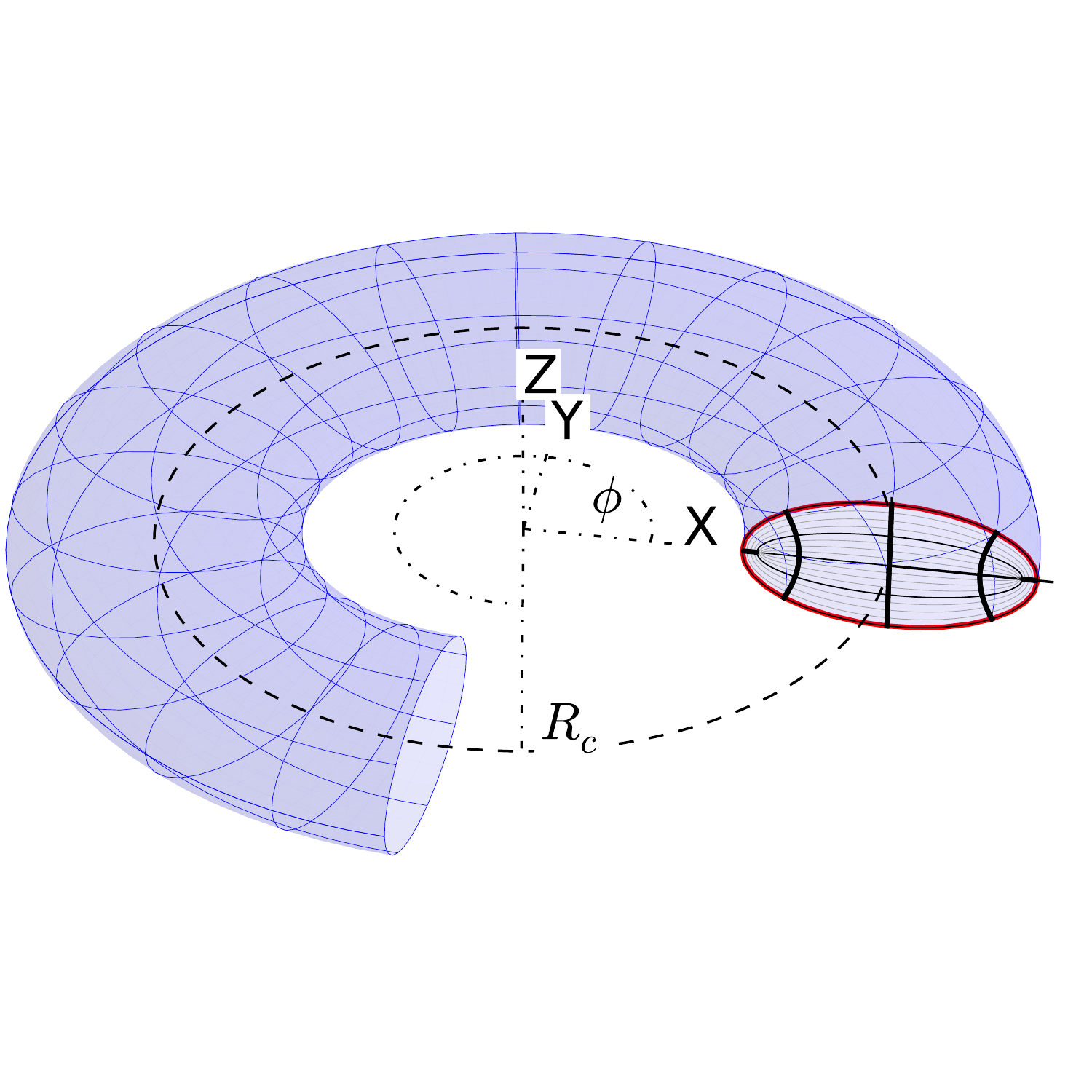}&
    \includegraphics[width=.5\columnwidth]{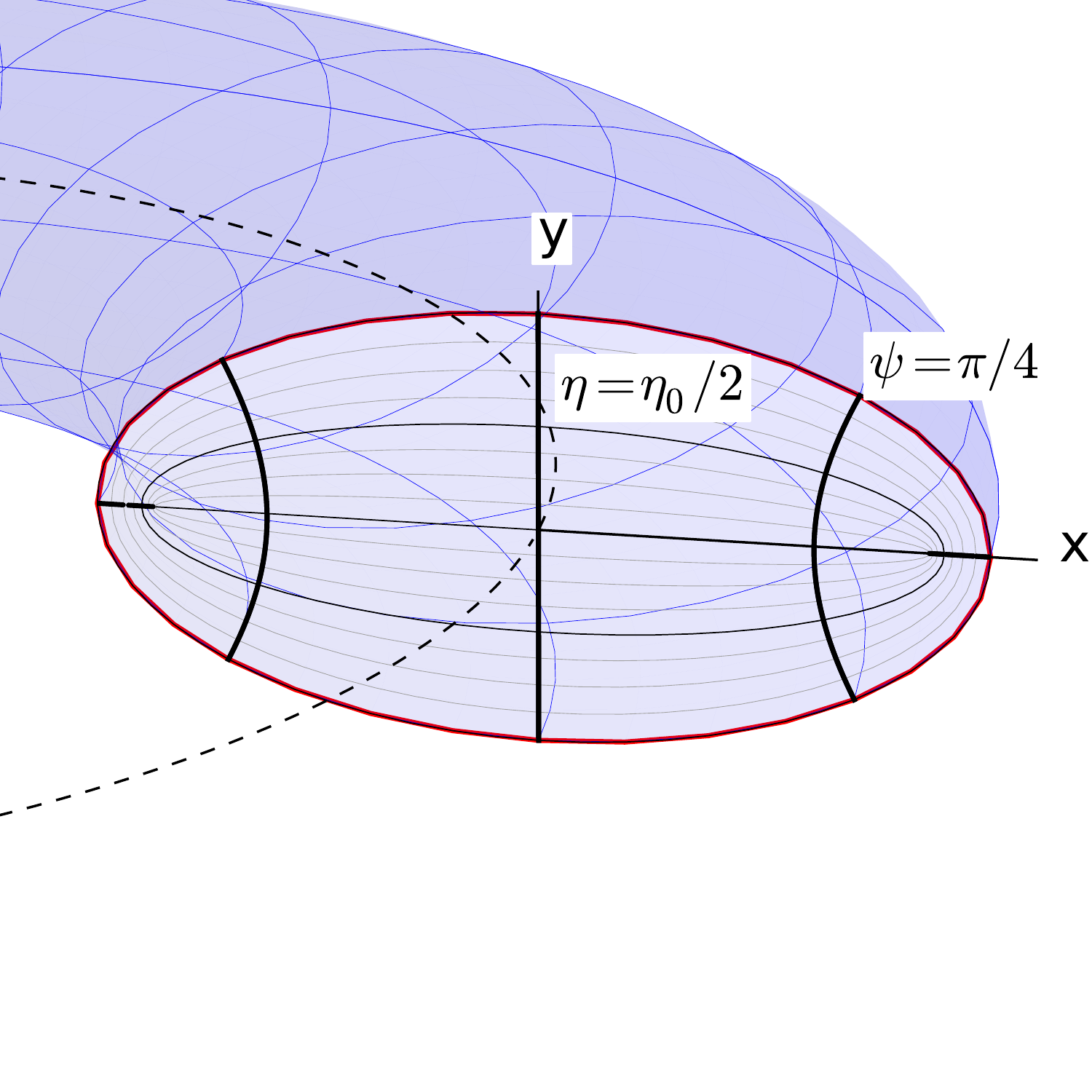}\\
  \end{tabular}
  \caption{The local elliptic toroidal coordinates. The left image
    gives the total torus while the right one shows the local
    coordinate system.}
  \label{fig:coor-local-elliptic-toroidal}
\end{figure}

Its transform to Cartesian coordinates is given by
(\cite{pbep:igte2010,pbep:mt22} see also Fig.~\ref{fig:coor-local-elliptic-toroidal})
\begin{eqnarray}
X &=&  \left(\rc \ + \ \be \ \cosh\bae \ \cos\baps \right) \cos\phi , \nonumber\\
Y &=&  \left(\rc \ + \ \be \ \cosh\bae \ \cos\baps \right) \sin\phi , \\
Z &=&   \ \be \ \sinh\bae \ \sin\baps , \nonumber
\end{eqnarray}
with $\be$ the eccentricity of the ellipse, $\bae$ and $\baps$ the
coordinates of the ellipse equivalent to $\eta$ and $\psi$ in
Eq.~\eqref{eq:elliptic-coordinates}.
The eccentricity of the ellipse is equivalent to the one for the
elliptic cylinder coordinates [Eq.~\eqref{eq:coor-ellipse-eccentricity}]
\begin{equation}
  \be = \sqrt{a^2 - b^2}
\end{equation}
with a the major and b the minor axes of the ellipse. The major radius of
the torus is $\rc$.
The boundary of the volume is now an ellipse instead of a circle. Its
surface is defined by  $\bar{\eta}_0$
\begin{equation}
  \tanh\bar{\eta}_0 = \frac{b}{a} .
\end{equation}
The volume of interest is a segment of a torus given by
\begin{equation}
0 \, \leq \, \bae \, \leq \, \bae_0, 
\qquad 
-\pi \, \leq \, \baps \, \leq \, \pi, 
\qquad 
- \phi_0 \, \leq \, \phi \, \leq \, \phi_0 .
\end{equation}

Equivalent to the ratio of minor to major radius $\eps$ one now
defines
\begin{equation}
\beps \ := \  \frac{\be}{\rc},
\end{equation}
thus  the ratio of the eccentricity over the major radius.
The  metric coefficients are then defined by
\begin{eqnarray}
h_t  &=& h_{\bae}  =  \bh_{\baps}  =  \be  \sqrt{\cosh(2 \bae) - \cos(2 \baps)}/\sqrt{2}\,, \\
h_\phi     &=& \rc  \  + \  \be \ \cosh\bae \cos\baps  \ \\
           &=& \ \rc \ (1 +  \beps   \ \cosh\bae \ \cos\baps ) \ = \ \rc \ \bh\,.
\end{eqnarray}
Up to some unessential constant factor the Laplace operator for toroidally (azimuthally) uniform potentials is
\begin{equation}
  \label{EllTorPotEq}
\begin{array}{l}
  \frac{1}{\cosh(2 \bae) - \cos(2 \baps)} \times  \bigg[ \frac{\pt^2}{\pt \bae^2}    +   \frac{\pt^2}{\pt \baps^2} - \\
 \ \ -  \frac{ \beps}{\bh} \left( \sinh\bae  \cos\baps  \frac{\pt}{\pt \bae}   +   \cosh\bae \sin\baps \frac{\pt}{\pt \bae} \right)\bigg] \bPh \ = \ 0.    
\end{array}
\end{equation}
The same approach is used as above for the local toroidal multipoles
(see section~\ref{sec:toroidal-circular-multipoles}):
 ``bending the basis functions'' is accomplished by replacing $\Phi$ with $\sqrt{\bh} \,\Phi$.
This yields
 \begin{eqnarray}
&&\frac{1}{\sqrt{\bh} }\times \\
\nonumber
&&\times \left[  \frac{\pt^2}{\pt \bae^2}   +    \frac{\pt^2}{\pt \baps^2}  -  \frac{\beps^2}{8 \bh^2}
\left( \cosh(2 \bae) - \cos(2 \baps) \right) \right] \left( \sqrt{\bh}  \bPh \right) =  0.
\end{eqnarray}
As above any terms of order $\order\beps 2$ are neglected. The remaining part of the differential equation then
resembles that for the elliptic cylindric multipoles (see section~\ref{sec:elliptic-multipoles}). Suitable solutions
of the latter equation must be identified with $ \sqrt{\bh} \bPh$.  Trials to do that in the same way as in the
elliptic cylindrical case by using the cylindrical solutions for normal or skew elliptic cylindrical multipoles
gave no satisfactory result. It is a task for the future to find suitable solutions.

\section{Measuring advanced multipoles}


The advanced multipole descriptions above are an extension to the 
standard description. Even if per se useful and furthering our unterstanding 
of magnetic fields in accelerators, these remain theoretical studies. In this chapter 
we show how these multipoles can be measured with rotating coil probes. The approach, 
the necessary measurement considerations and drawn conclusion are not limited to rotating 
coil probes but applicable to any device covering the curve of development similarily.

\subsection{Excurs: Rotating coil probes}

Rotating coil probes have been frequently used for measuring 
magnetic fields on a straight cylinder. In this paper 
only ``radial'' coil probes (see e.g. \cite{jain98:_harmon_coils,Davies:rotating_coils})
will be considered (see also Fig.~\ref{fig:rotating-coil-within-torus}). 
One can show that their 
 sensitivity $\cplx{K_n}$ of a so called radial coil probe is then
defined by
\begin{equation}
  \label{eq:rotating-coil-sensitivity}
  \cplx{K_n} = \frac{N L \rref}{n}
  \left[
    \left(\frac{r_2}{\rref}\right)^{n} -
    \left(\frac{r_1}{\rref}\right)^{n}
  \right]
\end{equation}
with $L$ the length of the coil probe, N its number of turns and $r_2$
the outer and $r_1$ the inner radius. Recording the induced signal,
calculating its spectrum and scaling the spectrum with the 
senstivity factors given above the harmonic content of the measured 
signal is obtained (e.g. \cite{jain98:_harmon_coils}).

\subsection{Measuring elliptic cylindric multipoles}
\label{sec:elliptic-multipoles-measurements}

While one could consider measuring the field along the elliptic
boundary and deducing harmonics from these measurements such
measurements are impractical within an anticryostat, which is only of
limited use as mechanical reference. Further these movements have to
be made with an accuracy of several $\mu$m and are thus not simply
applicable. Therefore a different approach is chosen here: the field
is measured at different lateral positions using a rotating coil probe
and the fields are then combined. A full description is given in
\cite{pbep:NIMA}, here only a short summary is given; it focuses 
on explaining the idea behind the method. This approach is not a 
mathematical rigorous optimisation; it uses the features of 
rotating coil probes, in particular the ones of a compensation array.
To the readers not familiar to rotating coil probes: the measurement
presented in the following is applicable to any system, which gives
more accurate results for the higher order harmonics than for the 
main one. This property implies that within the measurement area
the field homogeneity is obtained more precisely than the absolute value.
This property is used in the following.

\begin{figure}
  \centering
  \includegraphics[width=\columnwidth]{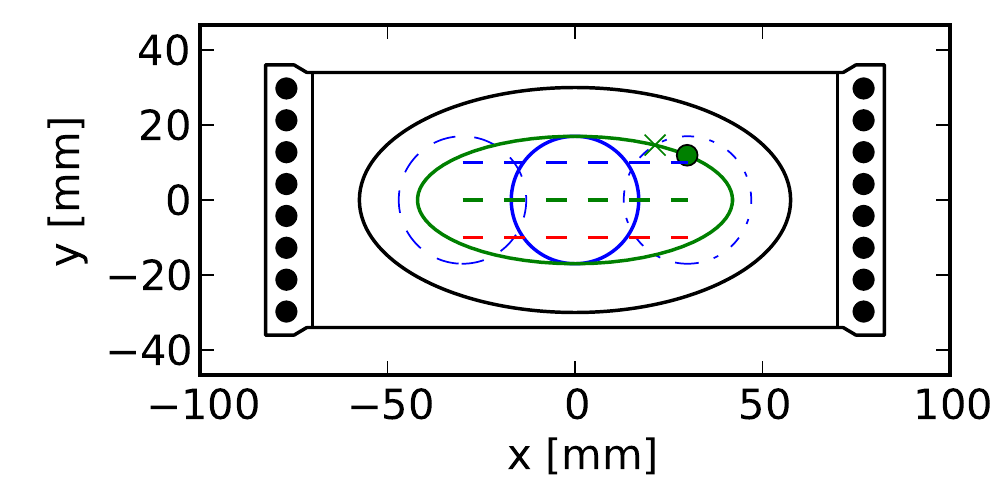}\\
  \caption{Location of the coil probes (blue circles, solid line for the centre,
    dashed lines for the left and dashed dotted for the right measurement) within the
    magnet aperture together with the ellipse (in green) used for
    calculating the elliptic multipoles. The larger black ellipse
    gives the intended good field region. The straight dashed lines
    show the area covered by the mapper and its hall probe. The other
    lines depict the magnet aperture together with the 8 turn coil
    (each turn shown as black circle). The intersection between the ellipse and the 
  circle are given by 'x' (with elliptic angle $\psi_c$. The green circle 
indicates $0.75\cdot\psi_c$.}
  \label{fig:sis100-dipole-coverage}
\end{figure}

\begin{figure}
  \centering
  \begin{tabular}[c]{@{\hspace{0em}}c@{\hspace{0em}}c@{\hspace{0em}}}
    \subfigure[Raw]{%
      \label{fig:sis100-dipole-uncorrected}
      \includegraphics[width=.5\columnwidth]{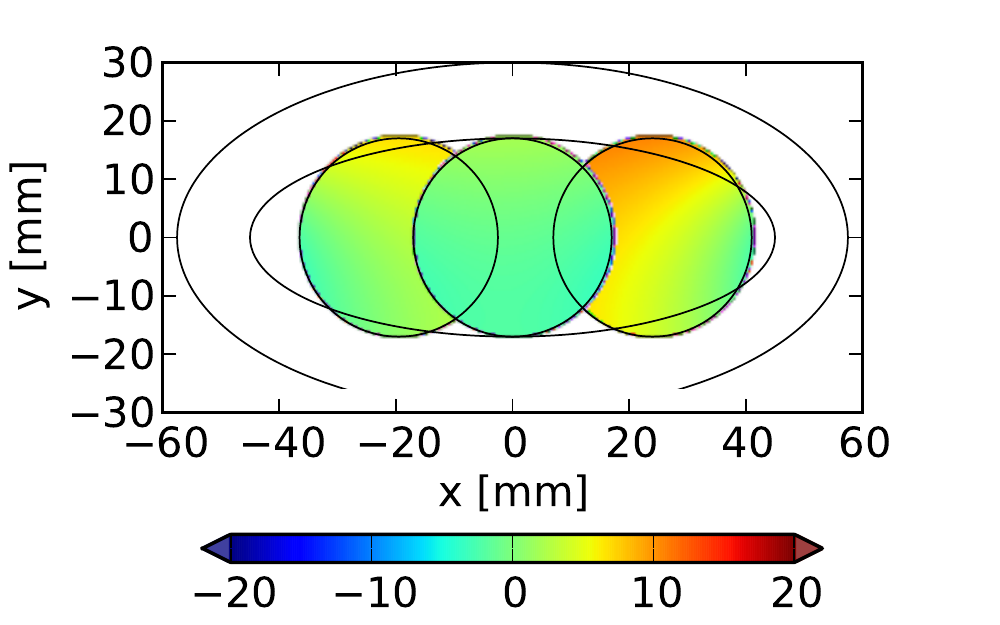}
    }%
    &
    \subfigure[corrected]{%
      \label{fig:sis100-dipole-corrected}
      \includegraphics[width=.5\columnwidth]{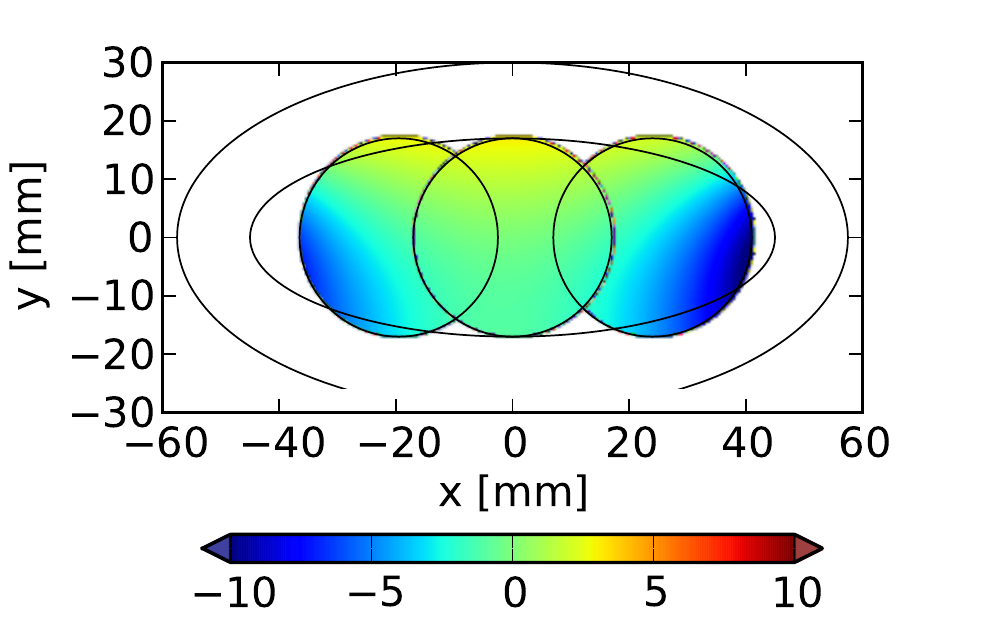}
    }%
    \\
  \end{tabular}
  \caption{Raw and  corrected data as measured by the rotating
    coil probe for the field homogeneity $\Delta B_y$ for a nominal
    field of 1.9~T.  Colour scale in units. 1 unit is equal to 100 ppm.  }
  \label{fig:sis100-dipole-corrected-uncorrected}
\end{figure}

The field is measured at different positions (see
Fig.~\ref{fig:sis100-dipole-coverage},
\cite{ipac14:elliptic_toroidal_applied,ipac14:sis100_fos_dipole})
with a rotating coil probe. If the field is then plotted within the
circles using the obtained multipole coefficients 
(see Fig.~\ref{fig:sis100-dipole-uncorrected}), 
one will find that the 
field is not continuous in the overlapping area, which is a clear violation 
of the potential equation or $\div B = 0$. Therefore the measurements have 
to be corrected. The source of this deviation is the limited accuracy of the 
main field measurement. Now one adjusts the main field (in magnitude and 
direction), so that the error in the overlapping error gets minimal.  
We have been using this approach for evaluating the 
measurements of the SIS100 dipole model magnets  and the SIS100 first series magnet
\cite{pac09:measure_calc,mole:mt21,ipac14:sis100_fos_dipole}.

\begin{figure}
  \centering
  \begin{tabular}[c]{@{\hspace{0em}}c@{\hspace{0em}}c@{\hspace{0em}}}
    \subfigure[raw data, By]{%
      \label{subfig:sis100-dipole-along-ellipse-raw-by}
      \includegraphics[width=.45\columnwidth]{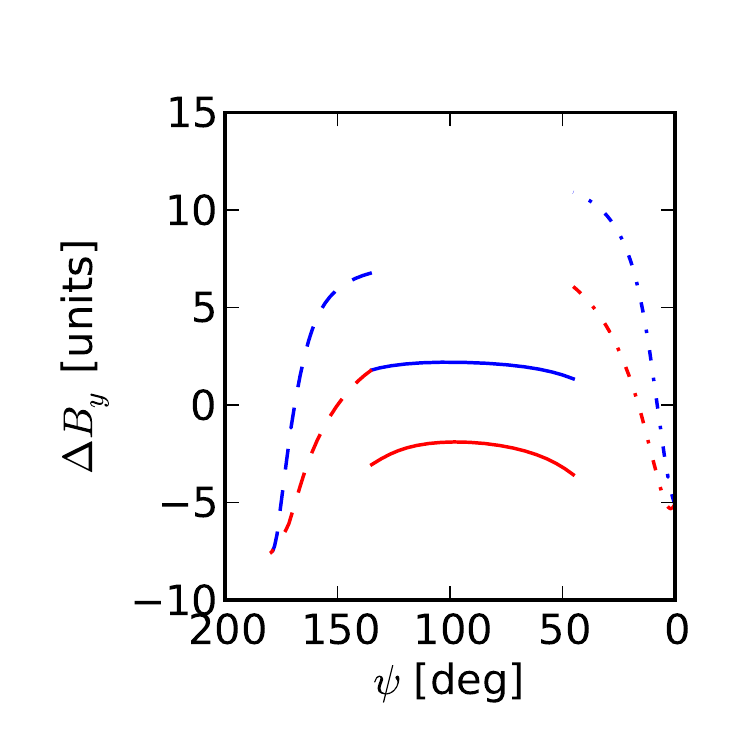}
      }
      &
    \subfigure[raw data, Bx]{%
      \label{subfig:sis100-dipole-along-ellipse-raw-bx}
      \includegraphics[width=.45\columnwidth]{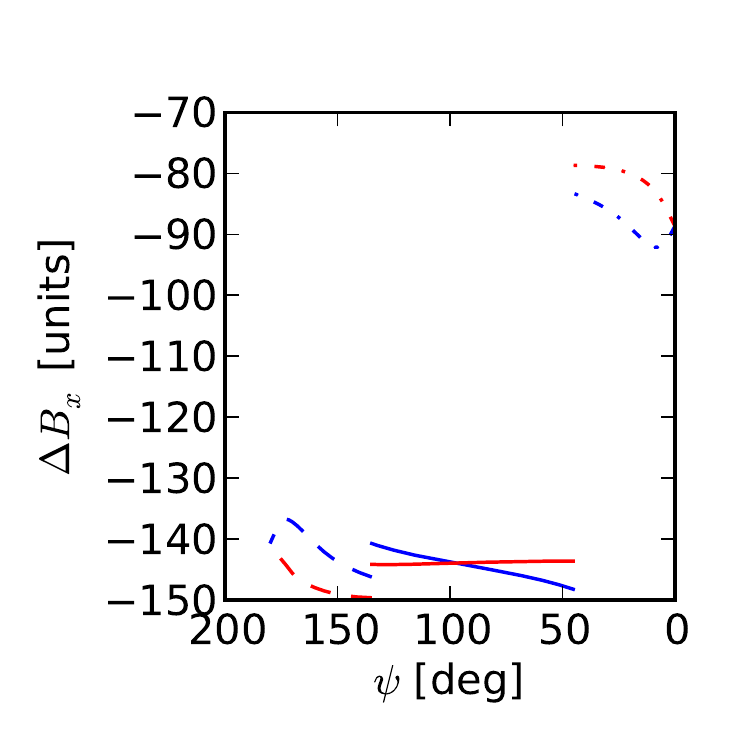}
      }
      \\
    \subfigure[corrected data, By]{%
      \label{subfig:sis100-dipole-along-ellipse-corrected-by}
      \includegraphics[width=.45\columnwidth]{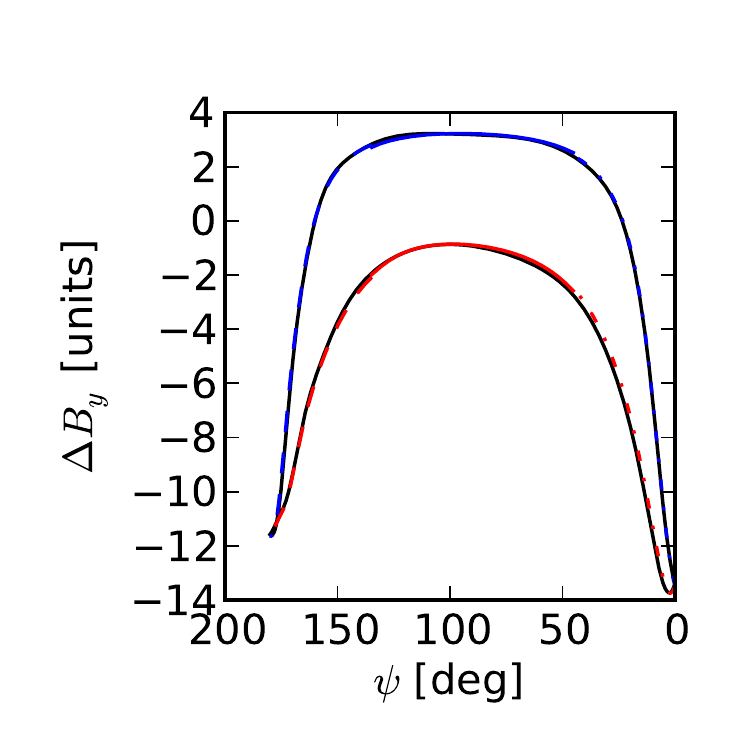}
      }
      &
    \subfigure[corrected data, Bx]{%
      \label{subfig:sis100-dipole-along-ellipse-corrected-bx}
      \includegraphics[width=.45\columnwidth]{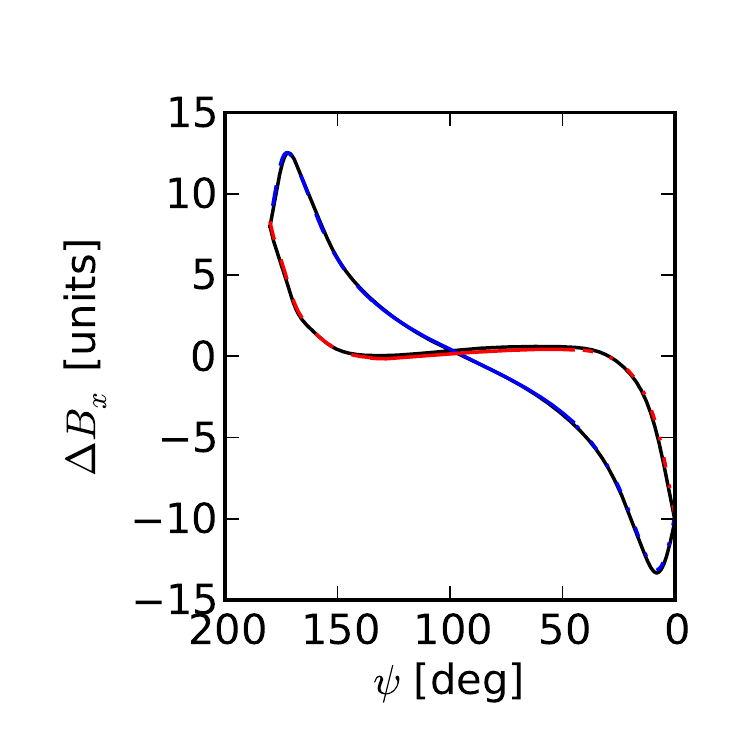}
      }
      \\
  \end{tabular}
  \caption{Raw and  corrected data as measured by the rotating
    coil probe for the field homogeneity $\Delta B_y$ for a nominal
    field of 1.9~T.  All data is plotted along the ellipse versus the angle $\psi$. For 
    angles from $0$ to $\pi$ the data are represented by blue lines for 
    angles from $-\pi$ to $0$ the data are represented by red lines. The solid lines 
    represent the data where the multipoles of the central coil probe were used. 
    The dashed lines where the multipoles of the measurement from the left side were used, 
    and the dashed dotted lines where the multipoles of the measurement from the right side 
    were used.
    For the left side the main multipole strength was corrected by $\approx 5$ units and 
    its angle by 15 mrad. For the right side the multipole strength was corrected by 
    $\approx -8$ units and its angle by $\approx$ 8 mrad. The black lines gives the 
    field when the derived multipoles are used for reinterpolation.
  }
  \label{fig:sis100-dipole-corrected-uncorrected-along-ellipse}
\end{figure}

To apply this method one can plot the data along the ellipse for the 
field $B_y$ and $B_x$ along the ellipse (see Fig.~\ref{subfig:sis100-dipole-along-ellipse-raw-by}
and Fig.~\ref{subfig:sis100-dipole-along-ellipse-raw-bx})
for the different measurements along the path one will use later for reconstructing the field 
on the ellipse (the choice of length is given below). One can see that the field 
one the upper path and the lower path are shifted by more or less the same value and that, despite
of the discontinuity, the curvature of the lines seems to be continuous.
 Now
the absolute strength of only the main multipole 
 and the angle of the main multipoles is corrected for the measurement at the left and the right
 until the lines are continuous. These parameters were adjusted manually using a computer program
with graphical interface.
The adaption of the strength of the main harmonic mainly affects $B_y$ and the 
correction of the angle of the main harmonics mainly affects $B_x$. 
The data interpolated from each measurement and the data reconstructed from the 
elliptic multipoles are presented in 
Fig.~\ref{subfig:sis100-dipole-along-ellipse-corrected-by}
and Fig.~\ref{subfig:sis100-dipole-along-ellipse-corrected-bx}. The data represented by the 
solid blue and red line represent the measurement of the centre circle (as shown 
by the solid blue circle in Fig.~\ref{fig:sis100-dipole-coverage}). The x-scale
 of the 
sub-figures in 
Fig.~\ref{fig:sis100-dipole-corrected-uncorrected-along-ellipse} was plotted from right to left.
It represents  the angle $\psi$ and hence the data obtained by the left measurement (shown as dash dotted line
in Fig.~\ref{fig:sis100-dipole-coverage} is also shown on the right in 
Fig.~\ref{fig:sis100-dipole-corrected-uncorrected-along-ellipse}).
One can see that the 
multipoles represent the original data significantly better than 0.1 units.

The last item that remains to clarify, is how one selects for which part of the ellipse
one uses which measurement data and where the cutting angle is chosen. Only the 
first quadrant is discussed here, as the others are treated accordingly. 
The simplest approach were, to use the data within the measurement on the right 
as long as the ellipse is within the area of this measurement and to use the
middle measurement for the rest in the quadrant. This approach is 
without physical justification. 
Further a discontinuity at the border of the measurement would be left
over. So a more sophisticated method is required.

\begin{figure}[tbp]
  \centering
  \includegraphics[width=\columnwidth]{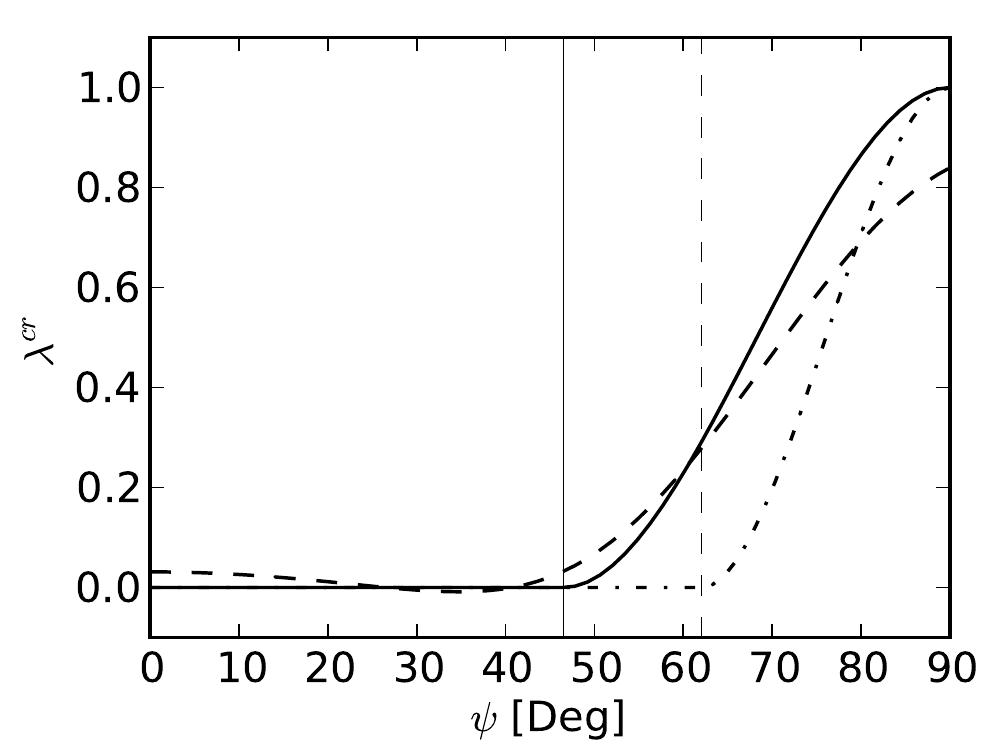}
  \caption{The weights of the measurement as well as the interpolation
    functions versus the angle $\psi$ for the first quadrant.
    The solid line indicates the
    function $\lambda$ with $p_0 = 0.75 \psi_c$ and the dashed dotted line the 
    function $\lambda$ with $p_0 =  \psi_c$.
    The dashed line indicates the weight
    function for the measurement in the centre  $2  \, \lambda^{cr} -
    0.5$. The scale and offset are used to facilitate the visual comparison
    to the function $\lambda$.    
    The vertical dashed line
    indicates the angle $\psi_c$      and the vertical solid line the angle
    $p_0 = 0.75\,\psi_c$. }
  \label{fig:NIMA-circle-weights}
\end{figure}

As start point the accuracy of the rotating coil
probe measurement has to be estimated in space.  The 
 authors chose  the distance of the point in question from the centre of rotation 
as weight function \cite{pbep:NIMA}.
Then  the data for the different points could be reconstructed 
for the different points $B_i$ using 
\begin{eqnarray}
    \label{eq:multipoles-measurement-interpolation}
  \cplx{ B_i(z)} &=& 
  \lambda\, B_1\, e^{\imath \alpha}\, \sum_{n =\, 2}^{N}\cplx{c_n}
  \left(\frac{{\bf z}}{\rref}\right)^n \\
  \nonumber
    &&+ (1 - \lambda) \, B^{r}_1\, e^{\imath \alpha_{r}} \,  \sum_{n =\, 2}^{N} \cplx{c^{l,r}_n}\left(\frac{{\bf z} - x_m}{\rref}\right)^n
\end{eqnarray}
for $B^{r}$ and $\alpha_{r}$ along the ellipse $\eta_0$. 
$r$ indicate the right measurement and $x_m$ the
offset of the axis of rotation from the centre.
$N$ was chosen to 10 as all further measured multipoles were close to zero.
The weight functions of the two measurements 
\begin{equation}
  w^r = \frac{R_m}{| {\bf z} - x_m |}\qquad
  w^c = \frac{R_m}{| {\bf z} |} \qquad
\end{equation}
is now combined to 
\begin{equation}
  \lambda^{cr} = \frac{w^{c}}{ \left(w^{c} + w^{r}\right)}.
\end{equation}
This $\lambda^{cr}$  (see also Fig.~\ref{fig:NIMA-circle-weights} \cite{pbep:NIMA})
represents the weight functions of the two measurements for
each position; but it is neither $0$ nor $1$ and the boundaries. Further it 
is desirable that the field and its derivative are continuous.
A good approximation can be made for the first quadrant using the polynomial
\begin{equation}
  \nonumber
\lambda(p_0) = 0, \quad \lambda(p_1) = 1, \quad
\lambda'(p_0, p_1) = 0,  \quad
  \lambda(p) =  3 p^2 - 2 p^3
\end{equation}
with 
\begin{equation}
  p = 
  \left\{ 
    \begin{array}{l@{\quad}l}
      0                                 & \psi < p_0 \\
      \frac{2 \psi - \pi}{2  p_0 - \pi} & p_0 \leq \psi \leq \pi\,;  \\
    \end{array}
  \right. 
\end{equation}
see \cite{pbep:NIMA}. $p_0$ was chosen such that $\lambda$ 
resembles the weight function as closely as possible: i.e $p_0 = 0.75\,\psi_c$, with $\psi_c$ the 
cut angle at which the ellipse intersects the right measurement area.
This definition of $\lambda$ ensures that the interpolation is continuous in its function
and its derivative.

The elliptic multipoles $\cplx{E_n}$ are then obtained 
by Fourier transform 
 according to
\begin{equation}
   \label{ell:Fcoff}
 {\bf E}_{n} 
= \frac{1}{{\pi \cal B}_0} \ \int_{-\pi}^{\pi}  
 {\bf B}_0 \big({\bf z} = e \cosh(\eta_0 + \imath \psi) \big) \ \cos(n \ \psi) \ d\psi .
\end{equation}

\begin{figure}
  \centering
      \includegraphics[width=\columnwidth]{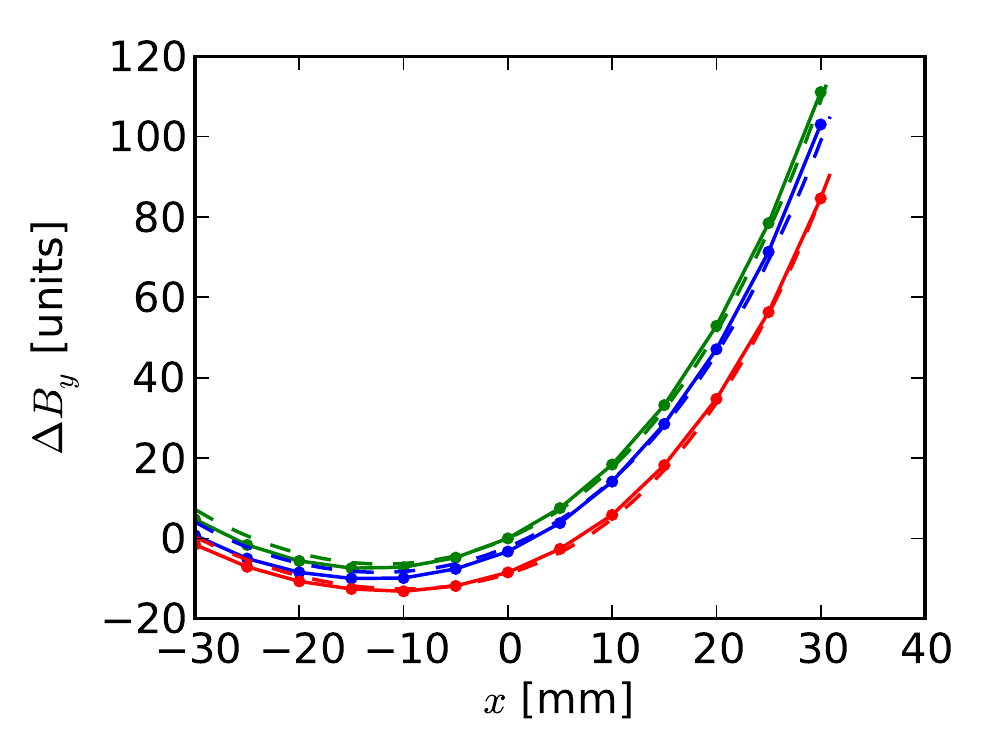}
      \caption{Comparison of the mapper data to the coil probe data. The
        dots indicate the mapper data, the dashed lines the reconstructed
    field for the magnet end. In
    blue the data are given for a measurement position y~=~+10~mm, in green for
      y~=~0 and in red for y~=~-10~mm. }
      \label{fig:sis100-coil-probe-mapper-end-field}
\end{figure}

The data obtained by these measurements were cross checked using a hall probe along a mapper and 
scanning the field along the lines presented in Fig.~\ref{fig:sis100-dipole-coverage}. 
The comparison of these two measurements is presented 
for the end field of the SIS100 FoS dipole in Fig.~\ref{fig:sis100-coil-probe-mapper-end-field}.
One can see that the two measurements match well. This proves that the 
method described here is sound and reliable and can be used for 
measuring accelerator magnets.


\subsection{Measuring toroidal  multipoles}
\label{sec:toroidal-multipoles-measurements}

The aperture of bending magnets (i.e. a dipole) in an accelerator can be reduced if
the magnet is curved and thus it follows the sagitta of the
beam. These magnets are typically measured with search coils,
i.e. coil probes which follow the magnets curvature. These were used
to measure SIS18 magnets \cite{sis18_magnet_measurement} the magnets of HIT
\cite{hit_magnets} or CNAO \cite{cnao_magnets_and_measurement}. Skew multipoles can not 
be derived from these measurements. Further the coil probe must be
aligned with the mid-plane. For magnets, whose yoke is operated at cryogenic temperature, 
search coils operating at cryogenic temperature were not seen as an option. 
An anticryostat with an reference surface was studied but not found to be a good technical 
option as it reduces the aperture too much when the SIS100 dipole magnets have to be 
measured. 

Therefore investigations were made if rotating coil probes can be used, as these 
can be operated within an anticryostat and allow measuring the integral 
harmonics content of the magnetic field.  So their 
theoretical measuring capabilities were studied for curved accelerator magnets 
using the local toroidal 
multipoles. 
\subsubsection{Coordinate systems}

\begin{figure}[b]
  \centering
  \includegraphics[width=.75\columnwidth]{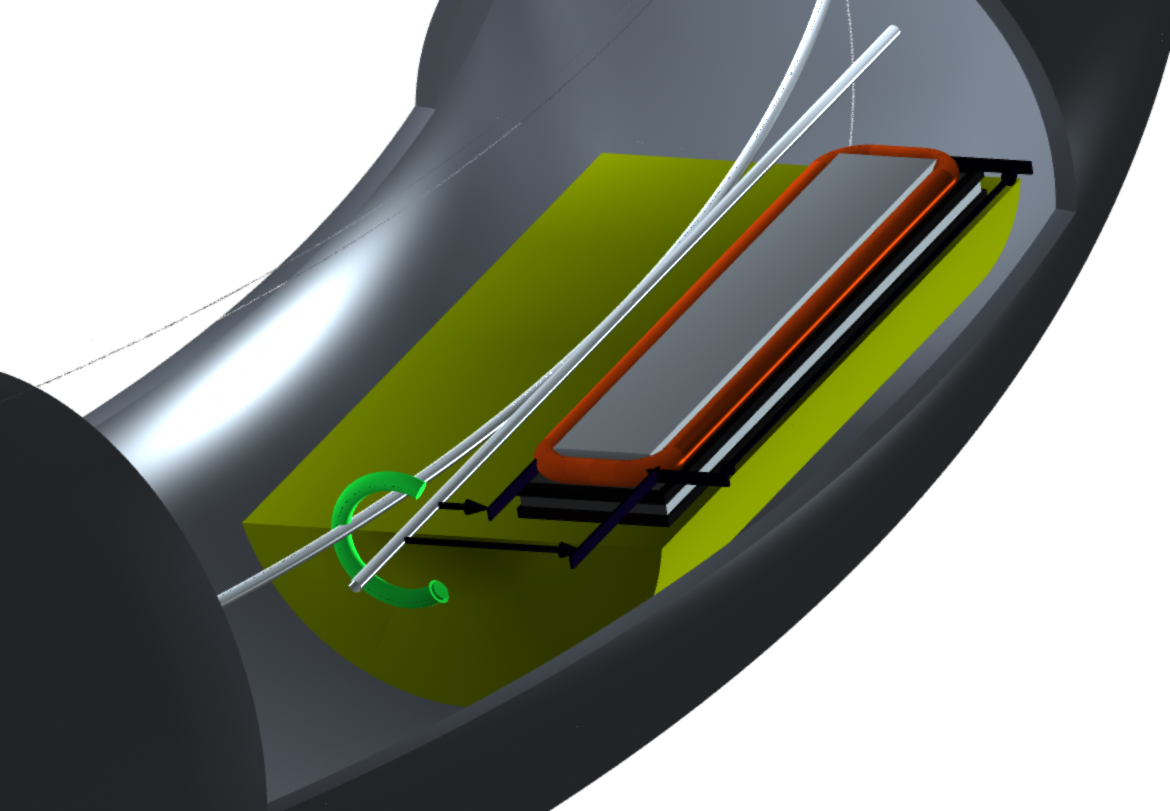}
  \caption{The rotating coil within a torus. The torus has been cut
    open so that the coil is visible. The centre line of the torus is
    indicated together with the rotation axis of the coil. The coil is
    indicated as a single turn. The rotation axis is a tangent to the
    torus. The centre of the rotating coil is also the point where
  the tangent touches the centre circle.}
  \label{fig:rotating-coil-within-torus}
\end{figure}

The toroidal circular multipoles (see section~\ref{sec:toroidal-circular-multipoles}) 
allow deriving the
limits of a rotating coil measurement. The results given here are 
based on \cite{pbep:compel,pbep:igte2010}.
The integrations are made similarly as for a straight magnet,
but now the dependence of the field 
along the direction of the rotation axis must 
be taken into account \cite{pbep:compel}.
Here one assumes that the field is constant
versus the toroidal angle $\phi$, with the  multipoles as given in Eq.~\eqref{Tor:BExp}
(see Fig.~\ref{fig:rotating-coil-within-torus}).
In the following only one half of the coil probe is considered (i.e. longitudinally from
the middle, where the coil axis touches the torus centre circle up to one end to avoid that 
some symmetric contributions would cancel).

Several coordinate systems are needed to deal with this complex
situation. The local Cartesian coordinates $x, y$ have been introduced in 
section~\ref{sec:toroidal-circular-multipoles}.
We assume that the axis of the rotating wire frame is parallel to the equatorial
plane. The centre of the frame has the coordinates  $x = d_x, \ y = d_y, \ \phi = 0.$
We introduce cylindrical coordinates $r, \vartheta, z$, whose
z-axis coincides with the rotation axis; the origin is at the centre of the wire frame.
The relation of the various local coordinates to the global Cartesian ones is given by 
\begin{eqnarray}
  \label{eq:measurement-toroidal-multipoles-x}
  X &=& (\rc + x ) \cos{\phi} = \rc + d_x + r \cos\vartheta\,,\\
  \label{eq:measurement-toroidal-multipoles-y}
  Y &=& (\rc + x ) \sin{\phi}  = -z\,,\\
  \label{eq:measurement-toroidal-multipoles-z}
  Z &=& y  =  d_y +  r \sin\vartheta\,.
\end{eqnarray}
Analytic relations between these various variables of these sets are needed for computing 
the  magnetic flux penetrating the coil probe. Some of these relations cannot be given 
exactly; again approximations up to the first order in $\eps$ are introduced. This is done 
for a fixed inclination angle $\vartheta$ of the coil's frame.  

\begin{figure}
  \centering
  \begin{tabular}[c]{@{\hspace{0em}}c@{\hspace{0em}}
    }
    \subfigure[front view]{%
      \includegraphics[width=\columnwidth]{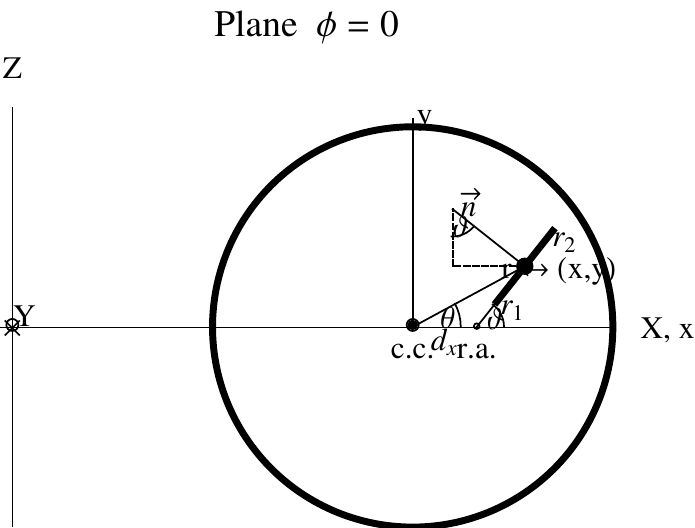}
      \label{subfig:measurement-toroidal-circular-coil-in-coordinates-front}
      }%
      \\
    \subfigure[top view]{%
      \includegraphics[width=\columnwidth]{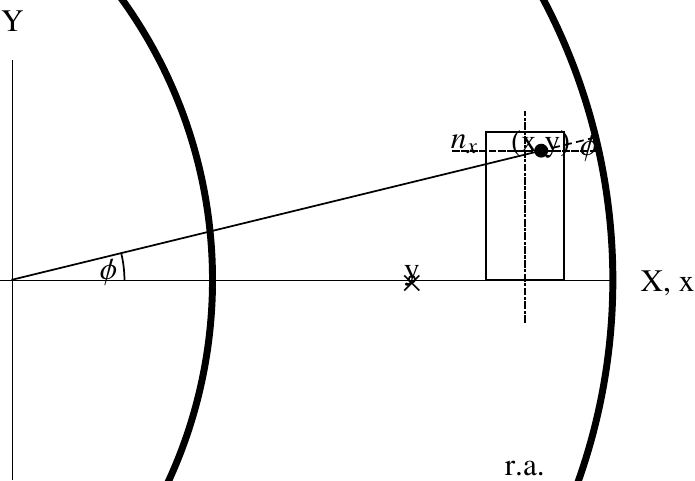}
      \label{subfig:measurement-toroidal-circular-coil-in-coordinates-top}
      }%
\\
  \end{tabular}
  \caption{Illustration of the coil and integration paths. The top
    figure shows the view from the front and the bottom the view from
    the top. $r_1$ and $r_2$ are the inner and outer radius of the
    coil probe, $d$ the offset of the coil probe rotation axis from the
    ideal axis, $z$ the longitudinal offset in the local coordinate
    frame $x,y,z$.}
  \label{fig:measurement-toroidal-circular-coil-in-coordinates}
\end{figure}

For that purpose one uses
the local Cartesian system of the toroidal multipoles 
(see section~\ref{sec:local-toroidal-coordinates} 
and Fig.~\ref{fig:measurement-toroidal-circular-coil-in-coordinates}).
From Eqs.~\eqref{eq:measurement-toroidal-multipoles-x} and  
\eqref{eq:measurement-toroidal-multipoles-y} follows
\newcommand{\xpos}{\rc + d_x + r \cos\vartheta}
\begin{equation}
  \tan\vph  =  \frac{Y}{X} = \frac{-z}{\xpos}
\end{equation}
the dependence of $\vph$ on the cylindrical variables. From this in turn we get:
\begin{eqnarray}
  \nonumber
 \cos\vph &=& \frac{1}{\sqrt{1 + \tan^2\vph}}  = \frac{\xpos}{\sqrt{\xpos + z^2}}.\\
\end{eqnarray}
Now we substitute $\rc = \rref/\eps$ and calculate the series which yields
\begin{equation}
  \label{eq:measurement-toroidal-multipoles-cos-phi}  
  \cos\vph  = \frac{-z}{\xpos} = 1 - \eps^2 \frac{z^2}{2 \rref^2} = 1  + \ordepsq.
\end{equation}

Inserting  Eq.~\eqref{eq:measurement-toroidal-multipoles-cos-phi}  
 in Eq.~\eqref{eq:measurement-toroidal-multipoles-x}
yields
\begin{eqnarray}
  x &=& \frac{\xpos}{\cos{\vph}} - \rc \\
\nonumber
&=& \sqrt{\left(\rc + d_x + r      \cos\vth\right)^2} - \rc \\
&=&\rc\left[1 +  \frac{d_x + \cos\vartheta}{\rc} + \frac{z^2}{2\rc^2} + \dots -1\right]\,.
 \end{eqnarray}
With these formulas the change in the $x$ position versus $z$ due to the
sagitta is given by 
\begin{equation}
  \label{eq:toroidal-measurement-x-vs-polar}
  x = \chi_x\left(r,\vartheta,z\right) := d_x + r \cos\vartheta + \eps \frac{z^2}{2\rref} + \ordepsq,
 \end{equation}
which approximates the larger torus circle with an parabola. 
According to Eq.~\eqref{eq:measurement-toroidal-multipoles-z}, $y$ is given by
\begin{equation}
  \label{eq:toroidal-measurement-y-vs-polar}
  y = \chi_y\left(r,\vartheta,z\right) := d_y + r \sin\vartheta\,.
 \end{equation}

 \subsubsection{The magnetic flux}

 The flux penetrating a coil probe (e.g. \cite{jain98:_harmon_coils})
 is given by
 \begin{equation}
   \label{eq:coil-in-torus-flux}
 {\Phi} \ = \ \int_{r_1}^{r_2} \int_{0}^{L}
   \mathbf{B}_n \ \ud z \ \ud r .
 \end{equation}
 $ \mathbf{B}_n$ is the magnetic induction normal to the wire frame.

 When calculating this quantity we must take into account that both
 the values of the local Cartesian variable change with $r, \vartheta,
 z$ as given in Eqs.~\eqref{eq:toroidal-measurement-x-vs-polar}
 and \eqref{eq:toroidal-measurement-y-vs-polar} and that the field
 components are transformed.  In a Cartesian system corresponding to
 the cylindrical system defined above, the normal is the unity vector
 $\vec{n}'$:
 \begin{eqnarray}
   \vec{n}' \ &=& \ (- \sin\vartheta',   \cos\vartheta', 0), 
   \nonumber
   \\
   \label{eq:toroidal-measurement-coil-area-vector-toroidal-angle}
   \vec{n}\ &=& \ (- \sin\vartheta \cos\phi,  \cos\vartheta, \ \sin\vartheta \sin\phi).
 \end{eqnarray}
 The angle $\vartheta$ gives the instantaneous inclination of the wire
 frame w.r.t. the equatorial plane defined by the X- and Y-axes (see
 Fig.~\ref{subfig:measurement-toroidal-circular-coil-in-coordinates-front}).  
\todo{Check that definition!}
The normal $\vec{n}$ is erected in the plane $\phi$ =  const in the local Cartesian system $x,y,z$.
It is assumed that the  
 $B_x$ and $B_y$ are invariant versus $\phi$.  $B_y$ is the
 same in this plane and in any other plane $z =$ const. 
In the
 transformation from the planes defined in the local Cartesian system 
 to the local toroidal  one the components $B_x$
 and the third component are involved. The latter is zero; the first
 one  contributes to $B'_x$ and $B'_z$ so we get (see
 Fig.~\ref{subfig:measurement-toroidal-circular-coil-in-coordinates-top})
 \begin{equation}
   \label{eq:toroidal-measurement-bfield-toroidal-angle}
   \vec{B}' \ = \ (B'_x, B'_y, B'_z) \ = \ (B_x  \cos\phi, B_y,  B_x  \sin\phi ).
 \end{equation}
 The coil probe is not sensitive to $B'_z$. The final expression for
 the component of the magnetic induction normal to the coil probe is:
 \begin{eqnarray}
   \label{eq:toroidal-measurement-normal-bfield-toroidal-angle}
   \mathbf{B}_n &=& \big(\vec{n}' \cdot \vec{B}' \big) \ = \ \big(\vec{n} \cdot \vec{B} \big) \ = \no  \\
   &=& -  B_x \big(\chi_x (r, \vartheta,z), \chi_y  (r, \vartheta,z) \big) \ \sin\vartheta \ \cos\phi + \no\\
   &&+ B_y \big(\chi_x  (r, \vartheta,z), \chi_y  (r, \vartheta,z) \big) \ \cos\vartheta . 
 \end{eqnarray}
 For the magnetic induction the expansion Eq.~\eqref{Tor:BExp} is inserted
into Eq.~\eqref{eq:toroidal-measurement-normal-bfield-toroidal-angle} 
with x [Eqs.~\eqref{eq:toroidal-measurement-x-vs-polar}] and y [Eq.~\eqref{eq:toroidal-measurement-y-vs-polar}]
replaced as just shown thus
 \begin{eqnarray}
   \label{eq:toroidal-measurement-normal-bfield-toroidal-angle-terms}
 \nonumber
 &&
 \cplx{B}_n  = \sum\limits_{m=1}^M \big[ \bar{B}_m \big(\cplx{T}_m^{Cn} (\chi_x, \chi_y)\cdot  \vec{n}\big)   
      +      \bar{A}_m \big(\cplx{T}_m^{Cs} (\chi_x, \chi_y)\cdot  \vec{n}\big) \big].\\
      &&
 \end{eqnarray}
The z-component of the surface normal to the coil area 
 [Eq.~\eqref{eq:toroidal-measurement-coil-area-vector-toroidal-angle}] can be set to zero 
as the coil probe is insensitive to  $B_z$.
 The integration is over $r$ and $z$ but 
takes  the
 curvature of the torus into account 
in the usual approximation (first order in
 $\eps$); so the flux seen by the coil with inclination $\vartheta$ is
 given by
 \begin{eqnarray}
   \label{eq:toroidal-measurement-flux-coil-integral}
   \Phi(\vartheta)  
   &=& \int_{0}^{L} \int_{r_1}^{r_2}  \mathbf{B}_n \ \ud r \ \ud z .
 \end{eqnarray}
 \todo{Following sentence correct?:}
  The
 integrals in this equation may be expressed in the following way
 after lengthy calculations performed with a computer algebra system
 (Mathematica\texttrademark):
 \begin{eqnarray}
   &&\int_{0}^{L} \int_{r_1}^{r_2}  \big(\cplx{T}_\mu^{Cn} (\chi_x, \chi_y) \cdot  \vec{n}\big) \ \ud r \ \ud z = \\
   && \quad = \frac{1}{N}
   \sum_{\nu=1}^{M+1} \left[   G_{\mu \nu}^{nc} \  K_\nu  \ \cos (\nu \vartheta)  +  
     G_{\mu \nu}^{ns} \  K_\nu  \ \sin (\nu \vartheta) \right] ,  \no \\
   && \hspace{.5\columnwidth}  \mu = 1, 2, ...,M ,\\
&&   \int_{0}^{L} \int_{r_1}^{r_2} \big(\cplx{T}_\mu^{Cs} (\chi_x, \chi_y) \cdot  \vec{n}\big) \ \ud r \ \ud z =\\
&& \quad= \frac{1}{N}   \sum_{\nu=1}^{M+1}  \left[  G_{\mu \nu}^{ss} \  K_\nu \ \sin (\nu \vartheta) +
     G_{\mu \nu}^{sc} \  K_\nu  \ \cos (\nu \vartheta) \right] .  \no 
 \end{eqnarray}  
 For the upper limit $M = 20$ is used; 
This gives sufficient accuracy
 for the applications we have in mind.  
simpler expressions 
could not be derived using the terms of  Eq.~\eqref{eq:local-toroidal-multipoles-basis-functions-solution1} separately.
Inspection of the elements of
 these four matrices show 1's, -1's respectively in the main diagonal
 of $G^{nc}, \ G^{ss}$ respectively.  All other elements are zero or
 their absolute values are appreciably smaller than unity. Therefore
 we define:
 \begin{eqnarray}
   G_{\mu \nu}^{nc} \ &=& \,\,\,\ \delta_{\mu \nu}  \ + \ H_{\mu \nu}^{nc} ,  \no \\
   G_{\mu \nu}^{ss} \ &=&         - \delta_{\mu \nu}  \ + \ H_{\mu \nu}^{ss} ,  \no \\
   && \hspace{.3\columnwidth}  \mu , \nu  = 1, 2, ...,M ,\\
   G_{\mu \nu}^{ns} \ &=& \ H_{\mu \nu}^{ns} ,  \no \\
   G_{\mu \nu}^{sc} \ &=& \ H_{\mu \nu}^{sc} . \no 
 \end{eqnarray}  
 The resulting $M \times M$ matrices are used to define a $2 M \times
 2 M$ matrix:
 \begin{equation}
   G :=  G^o  +  H \ = \ G^o  + 
   \left( 
     \begin{array}{cc}
       H^{nc} &   H^{ns}  \\
       H^{sc} &  H^{ss}   
     \end{array}
   \right)
 \end{equation}
 The $2 M \times 2 M $ matrix $G^o$ is a diagonal matrix comprising M
 1, then M -1's:
 \begin{equation}
   G^o =  dia(1,1,1,\dots,\dots,-1,-1,1) = (G^o)^{-1}.
 \end{equation}
 The elements in columns 2 to M of the $M \times M$ matrices are the
 same (or the opposite) at corresponding places:
 \begin{eqnarray}
   G^{nc} + G^{ss} &=& H^{nc} + H^{ss} = (H^r, 0,0,0,\dots) , \\
   G^{ns} -  G^{sc} &=& H^{ns} -  H^{sc} = (H^i, 0,0,0,\dots) ;   \\
   (H^r + \imath H^i)_k   &=& \epsilon  \cplx{dz}^k/(2k  \rref^k),  \cplx{dz} = dx + \imath dy . \quad\quad
 \end{eqnarray}
 In the brackets at the end of the first two lines of the above
 equations the first columns are listed.  Only the first column
 contains elements different from zero; which are given in the last
 line, the label $k$ denotes the k-th row of this first column.

 The flux $\Phi(\vartheta) $ may then be written as:
 \begin{equation}
   \Phi(\vartheta)  =  \frac{1}{N} \ (B, A) \cdot G \cdot K \cdot v
 \end{equation}
 where K is a diagonal matrix containing the M sensitivities
 twice:
 \begin{equation}
   K := dia(K_1, K_2, ..., K_M, K_1, K_2, ... , K_M);
 \end{equation}
 $(B, A)$ denotes a row vector comprising the M + M expansion
 coefficients:
 \begin{equation}
   (B, A) =  (B_1, B_2, B_3, \dots, B_M, A_1,  A_2, A_3, \dots, A_M);
 \end{equation}
 The column vector $v$ contains the M harmonics cos$(\nu \vartheta)$
 at first, then the M harmonics sin$(\nu \vartheta)$:
 \begin{eqnarray}
   v &= \big(& \cos( \vartheta), \cos(2 \vartheta), \dots , \cos( M \vartheta),\nonumber\\
   &&\sin( \vartheta), \sin(2 \vartheta), \dots , sin(M \vartheta) \big).
 \end{eqnarray}

 Integrating Faraday's law with respect to time
 \begin{equation}
   \int V(t) dt  = - \ N \ \int\int_{\cal F} \mathbf{B}_n \ \ud r \ \ud z .
 \end{equation}
 integrating a Fourier expansion of the Voltage induced in the coil
 \begin{equation}
   \int V(t) \ dt  \ = \ \sum_{n=1}^M  \left[ a_n \ \cos (n \omega t) \ + \ b_n \ \cos (n \omega t)  \right]
 \end{equation}
 and identifying $\vartheta = \omega t $ assuming that the coil
 rotates with constant angular velocity $\omega$ we finally get:
 \begin{eqnarray}
   (a, b) \cdot v  &=&  -  \ (B, A) \cdot G \cdot K \cdot v ,\\
   (B, A) &=& - \ (a, b) \cdot K^{-1} \cdot G^{-1} .
 \end{eqnarray}
 (a, b) denotes the row vector comprising the Fourier coefficients of
 the signal in the rotating radial coil:
 \begin{equation}
   (a, b) = (a_1, a_2, ... , a_8, b_1, b_2, .... , b_8).
 \end{equation}
 A good approximation for the inverse ot the matrix $G$ is given by :
 \begin{equation}
   G^{-1} \approx G^o -   G^o \cdot H \cdot G^o . 
 \end{equation}


 \subsubsection{Conversion matrices}

The sections given above showed that the toroidal multipoles can be deduced from rotating coil probe 
measurements assuming that these are uniform over the measurement length. The calculations above
also considered the effect of a misplaced coil. 
The multipoles are mapped  to the complex multipoles by
\begin{equation}
  \left(
  \begin{array}[c]{c}
    \cplx{T}_{\mu}^{Cn}\\
    \cplx{T}_{\mu}^{Cs}
  \end{array}
  \right)
  = 
  \left(
  \begin{array}[c]{c|c}
    G^{nc} & G^{ns} \\
    \hline
    G^{sc} &  G^{ss}\\
  \end{array}
  \right)
  \left(
  \begin{array}[c]{c}
    \vec{B_n}\\
    \vec{A_n}
  \end{array}
  \right)\, .
\end{equation}
Each of the submatrices    $G^{nc}$, $G^{ns}$,     $G^{sc}$ and   $G^{ss}$ 
  is set up by 
\begin{equation}
  G =  I + \upmat^{dr} + \epsilon \left(\upmat^{L} + \mat{U} + \upmat^{sk} + \upmat^{{R2}_0} + \upmat^{R2} \right).
\end{equation}
Only elements of $\upmat^{L}$ depend on the coil length $L$ 
while only
elements of $\upmat^{sk}$ depend on the coil sensitivity parameters [Eq.~\eqref{eq:rotating-coil-sensitivity}]. The elements of the 
last matrix depend only on $d_x$ and $d_y$. All these matrices can be derived from complex matrices, but the
 result itself is not analytic. 
\todo{Stimmt das?} In the following part the coefficients of the different submatrices will be given.

\newcommand{\trow}{n}
\newcommand{\tcol}{m}

The matrix $\mat{U}$ is the sole one which consists of constant terms and is given 
by 
\begin{equation}
  \mat U = \frac{\trow + 1}{4 \trow} \kroneckerdelta_{\trow+1,\tcol} \,. 
\end{equation}
Many of the matrices below are given as triangular lower
matrices. Therefore one defines
\begin{equation}
  \upmat_{\trow,\tcol} =
  \begin{cases}    
        1 & \trow \ge \tcol\\
        0     & \trow < \tcol\, . \\
  \end{cases}
\end{equation}
Similar to measuring with rotating coil probes, an offset of the coil probe causes that one multipole creates spurious
other multipoles. These are similar for the different submatrices of matrix $H$ and thus summarised here.
The matrix $\upmat^{dr}$ is the only one, which does not depend on the 
torus curvature ratio $\epsilon$. Its non zero elements are given by
\begin{equation}
\label{eq:feed-down}
\upmat_{\trow,\tcol}^{dr}=  \binom{\trow - 1}{\tcol - 1} \left(\frac {d_x + \imath d_y} \rref\right)^{\trow - \tcol}  *\upmat_{\trow,\tcol} - I\,.
\end{equation}
The ``*'' denotes that these multiplication is to be made element wise.
This term is due to the frame translation in $d_x$ and $d_y$, which is exactly the same as found if a rotating coil probe
is displaced by  $d_x + \imath d_y$ within a cylindric circular coordinated system. This effect is  called  the ``feed-down'' effect.
The identity matrix is subtracted as the diagonal has to be singled out for later treatment.

For describing $\upmat^{dr2}$ one defines 
\begin{equation}
  \upmat^{\ud}_{\trow,\tcol}=
  \begin{cases}    
        1 & \trow > \tcol + 1\\
        0     & \trow \le \tcol + 1\,.
  \end{cases}
\end{equation}
Then $\upmat^{dr2}$ is given by
\begin{eqnarray}
\label{eq:feed-down-curvature}
\nonumber
\upmat_{\trow,\tcol}^{dr2} &=&  
(\trow - \tcol) \binom{\trow - 1}{\tcol -1}
\left(\frac {d_x +    \imath d_y} \rref\right)^{\trow - \tcol - 1} * \upmat^{\ud}_{\trow,\tcol} + \\
&& + \tcol\    \kroneckerdelta_{\trow, \tcol+1}  = \rref  \frac{\ud}{\ud z} \upmat^{dr}
\end{eqnarray}
and is similar to Eq.~\eqref{eq:feed-down} except for the binomial factor and that the power is reduced by 1. 
The dependence on $L$ is given by
\begin{equation}
  \upmat_{\trow,\tcol}^{L} = \frac{L^2}{3 \rref^2} \upmat^{dr2} \,.
\end{equation}
Please note, that the first side band includes the
  constant term $L^2 / (3 \rref^2)$.
The dependence on the coil sensitivity factors $\cplx{K_n}$ is given by 
\begin{equation}
  \label{eq:measuring-toroidal-coil-sensitivity-influence}
  \upmat_{\trow,\tcol}^{sk}= \frac{1}{4\, \left(\tcol+1\right)} * \frac{\cplx{K_{\tcol+2}}}{\cplx{K_\tcol}} * \upmat^{dr2} \,.
\end{equation}
 The most complex matrix does only depend on $d_x$ and $d_y$. It
 is
given by 
%
\begin{eqnarray}
  \upmat_{\trow,\tcol}^{R2}&& = 
\left(\frac{\trow\, \tcol}{\trow - \tcol + 1} * \upmat_{\trow,\tcol}  +   \kroneckerdelta_{\tcol, 1}\right) * \\
\nonumber
&&
*    \left[\frac{d_y}{\rref} -  \left(\frac{2 - \tcol + 2 \trow}{\tcol} * \upmat_{\trow,\tcol}  - \trow      \kroneckerdelta_{\tcol, 1}  \right)\imath \frac{d_x}{\rref}\right] *\\
\nonumber
&&
*  \frac{1}{4 \trow} * \left(\upmat^{dr} + I  \right)\,.
\end{eqnarray}
The last matrix $\upmat^{R2_0}$ is given by
\begin{equation}
\upmat^{R2_0}_{\trow,\tcol} =\frac{1}{2  \trow} \left(\frac{d_x + \imath d_y}{\rref}\right)^{\trow}\kroneckerdelta_{\tcol,1}.
\end{equation}

Each of the submatrices    $G^{nc}$, $G^{ns}$,     $G^{sc}$ and   $G^{ss}$ 
  is set up by 
\newlength{\theIlength}
\settowidth{\theIlength}{$ +I $}
  \begin{equation}
\label{eq:measuring-toroidal-circular-conversion-matrix}
  \begin{array}[c]{lcll@{\hspace{.1em plus .1em}}l@{\hspace{.1em plus .1em}}l}
  G^{nc} &=&   I &  +  \re\left[\upmat^{dr}\right] + 
 \epsilon
\big(
    &- U &
    + \re\left[\upmat^{L}\right] 
    - \re{\left[\upmat^{sk}\right]}
\\ 
  &&&&&  
+\im{\left[\upmat^{R2}\right]}
    + \re{\left[\upmat^{{R2}_0}\right]} 
  \big),\\
  G^{ns} &=&    &- \im\left[\upmat^{dr}\right] +  
  \epsilon
  \big(
    & &
    - \im\left[\upmat^{L}\right]+ \im\left[\upmat^{sk}\right]
\\
      &&&&&
    +\re{\left[\upmat^{R2}\right]}
  \big),\\
  G^{sc} &=&     &- \im\left[\upmat^{dr}\right] + 
  \epsilon
  \big(
  & &
    - \im\left[\upmat^{L}\right]
    + \im\left[\upmat^{sk}\right]
\\
       &&&&&
+ \re{\left[\upmat^{R2}\right]}
    - \im{\left[\upmat^{{R2}_0}\right]}
  \big),\\
  G^{ss} &=&   -I& - \re\left[\upmat^{dr}\right] + 
\epsilon
  \big(
    &+ U&
    -\re\left[\upmat^{L}\right]+\re{\left[\upmat^{sk}\right]}   
\\
       &&&&&
    - \im{\left[\upmat^{R2}\right]}
  \big).      
  \end{array}
\end{equation}
This summary already shows that  the main dipole is affected by all measured harmonics.
On the other hand $\epsilon$ is rather small for the machines used here. The different matrices $H$
are obtained ommitting the identity matrices. Comparing 
the operators $\re$ and $\im$ on $\upmat^L$ and $\upmat^{sk}$  to the ones operating on 
$\upmat^{dr}$ in Eq.~\eqref{eq:measuring-toroidal-circular-conversion-matrix} one can assume that
 $\upmat^L$ and $-\upmat^{sk}$ are analytic.

\subsubsection{Choosing a coil probe length}
\label{sec:measuring-local-toroidal-circular-coil-probe-length}
Evaluating all the different terms one can see that only $\upmat^{L}$ is of significant size for 
accelerator magnets with characteristic values as given in 
Table~\ref{tab:toroidal-circular-multipoles-coefficients-different-machines}.
\begin{table}
  \centering
  \caption{Parameters for different machines. }
  \label{tab:different-machines}
  \label{tab:toroidal-circular-multipoles-coefficients-different-machines}
  \begin{ruledtabular}    
  \begin{tabular}[c]{lrrrrr}
          & \multicolumn{1}{c}{$\rc$ [m]}
          & \multicolumn{1}{c}{$\rr$ [mm]} 
          & \multicolumn{1}{c}{$\epsilon$ [units]} 
          & \multicolumn{1}{c}{$L$ [mm]} 
          & \multicolumn{1}{c}{$d_x$,$d_y$ [mm]}
          \\
          \hline
  LHC     & 2804 & 17 &  0.04& 600 &1\\ %
  SIS100  & 52.5 & 40 & 7.62& 600 &1\\
  SIS300  & 52.5 & 35 & 6.67& 600 &1\\
  NICA    & 15   & 40 & 26.67& 600 &1\\
  \end{tabular}
  \end{ruledtabular}
\end{table}
 A criterion can be given for defining an adequate coil probe length by 
demanding that the feed down effect as found for cylindric circular multipoles and for 
measuring toroidal circular multipoles should be of equivalent size. 

For describing the relation one defines $\upmat^{\ud L}_{\trow,\tcol}$
\begin{equation}
  \upmat^{\ud L}_{\trow,\tcol}=
  \begin{cases}
        1 & \trow > \tcol\\
        0     & \trow \le \tcol + 1\,.
  \end{cases}
\end{equation}
Then the relation can be given by 
\begin{equation}
  \upmat_{\trow,\tcol}^{dr} =  \underbrace{\frac{3 \rref^2}{\eps L^2} \frac{d_x + \imath d_y}{\rref}}_{L_s} \left(\frac{1}{\trow - \tcol}\upmat^{\ud L}_{\trow,\tcol}\right) * \upmat_{\trow,\tcol}^{L} \,. 
\end{equation}
Demanding that the feed down effect due to coordinate translation $\upmat^{dr}$ shall be 
similar to  $\upmat^{L}$ then   $L_s = 1$  
which yields as relation for the coil length
\begin{equation}
\label{eq:toroidal-circular-multipoles-coil-length}
  L = \sqrt{\frac{3 \left(d_x + \imath d_y\right) \rref}{\epsilon}} =
  \sqrt{3 \left(d_x + \imath d_y\right) R_C},
\end{equation}
using only the first side band (n-m = 1). 
Higher harmonics will be affected by larger spurious harmonics due to $\upmat^{L}$.
As one can see the to be chosen coil length $L$ becomes larger
when the displacement errors $d_x$ and $d_y$ get smaller.
On the other hand a smaller $d_x$ or $d_y$ will create smaller total spurious harmonics, 
and thus the overall contribution gets small. Decreasing $d_x$ and $d_y$ by a 
magnitude will only allow increasing the coil length by a factor of 3.
It is recommended to 
evaluate the above equation for the maximum accepted deviations $d_x, d_y$. If a longer coil probe is chosen 
more effort shall be taken to determine $d_x$ and $d_y$ so that an appropriate treatment of 
the feed down effect can be made.

The
attention of readers familiar to coil probes and evaluating their measurements  shall be drawn to 
the influence of the sensitivity factors
[Eq.~\eqref{eq:measuring-toroidal-coil-sensitivity-influence}]. The
first term affecting the ``dipole'' is the ``sextupole'' term. Here
the ratio can be very small, if compensating systems or ``bucking''
systems are used (see e.g. \cite{jain98:_harmon_coils}). Any further
treatment will require to invert the matrices
Eq.~\eqref{eq:measuring-toroidal-circular-conversion-matrix}; in this case the sign of the term will 
swap and its magnetitude change should be rather small, given that the identity matrix is \todo{partly} involved.

\subsubsection{Magnitude of the terms}

The formulae given above were evaluated for the following different machines:
the Large Hadron Collider (LHC) at CERN\cite{lhc_design_report},
 SIS100 \cite{magnet:mt21,tdr} and
SIS300 at GSI, and NICA \cite{rupac:nica_magnets,nica:asc10} at Dubna
(see Table~\ref{tab:different-machines}). 
The parameters given in Table~\ref{tab:different-machines} were used to calculate
the coefficients of the matrices. 
Accelerators require a field description with an accuracy of 1 unit and
 roughly 0.1 unit for the field homogeneity (1 unit equals 100 ppm). 
Therefore any contribution less
than 1 ppm can be ignored.

Due to the circumference of the LHC $\epsilon$ is very small and thus the 
correction of all matrices are very small (less than 1 ppm) except for the 
matrix $\upmat^{L}$, where the values close to the diagonal get to a 
size of 2000 ppm for $d = \rref $. So even for an insane value of
$d_z$ the artefacts are handable.
This value may seem to exceed the target value for the field description;
but the higher order multipoles are in the order of 100 ppm; thus the effective
artifact will be safely below the target value of 10 ppm.

For machines with an aspect ratio as found for SIS100 or SIS300 the matrix $\mat{U}$
is in the order of 100 ppm. It can be neglected except for the main multipole.
The values of the matrix  $\upmat^{L}$ get of similar size as the values 
for $\upmat^{dr}$. The magnitude of these values are defined by the magnitude of the offset
$\left|d_x + \imath d_y\right|$. 
Also when measuring straight magnets special methods are applied to 
obtain the offset $d$ from the measured dataset \cite{jain98:_harmon_coils}. 
Therefore one can assume that 
the artifacts can be minimised by similar adequate procedures. 

The parameters for the different machines are given in Table~\ref{tab:toroidal-circular-multipoles-coefficients-different-machines}. 
A practical coil length was deduced imposing that the influence of the offset of the coil from the straight line shall be
of the same order as for a coil probe measuring a straight magnet [Eq.~\ref{eq:toroidal-circular-multipoles-coil-length}]. 
Now the  matrices are  evaluated to see to which extend different toroidal multipoles correspond to 
one measured multipole $B_n$ or  $A_n$.

The matrix Eq.~\eqref{eq:measuring-toroidal-circular-conversion-matrix}
was evaluated. For the geometry of the machines considered here, as
listed in
Table~\ref{tab:toroidal-circular-multipoles-coefficients-different-machines},
only the  terms $U$ and $\upmat^{L}$ have a significant contribution;
thus only the expression $U$ +  $\upmat^{L}$  is evaluated below. It is given by
\begin{eqnarray}
\label{eq:toroidal-circular-multipoles-measurement-matrix}
\small
\mat{C}_1 &=& U + \upmat^{L} \\
\nonumber
&= &
\left(
\begin{array}{cccccc}
 \cdot & \frac{1}{2}           &                              &   \\
 \frac{L^2}{3 \rref^2}         & \cdot                        & \frac{3}{8} \\
 \frac{2 L^2 d_z}{3 \rref^3}   & \frac{2 L^2}{3 \rref^2}       & \cdot & \frac{1}{3}         \\
 \frac{L^2 d_z^2}{\rref^4}     & \frac{2 L^2 d_z}{\rref^3}      & \frac{L^2}{\rref^2}         & \cdot & \frac{5}{16}             \\
 \frac{4 L^2 d_z^3}{3 \rref^5} & \frac{4 L^2 d_z^2}{\rref^4}    & \frac{4 L^2 d_z}{\rref^3}    & \frac{4 L^2}{3 \rref^2}         &  \cdot & \frac{3}{10}  \\
 \frac{5 L^2 d_z^4}{3 \rref^6} & \frac{20 L^2 d_z^3}{3 \rref^5} & \frac{10 L^2 d_z^2}{\rref^4} & \frac{20 L^2 d_z}{3 \rref^3}    & \frac{5 L^2}{3 \rref^2} &  \cdot   \\
\end{array}
    \right)\,.
\end{eqnarray}
The dots indicate the diagonal, where all elements are zero. The parameters given in Table~\ref{tab:toroidal-circular-multipoles-coefficients-different-machines}
were inserted. The matrix $C$ was inverted which gives
\begin{equation}
\small
C^{-1} = I + \frac{1}{10 000} 
\left(
\begin{array}{rrrrr}
   . &   4 &   . &     &     \\
 143 &   . &   3 &     &     \\
   3 & 285 &   . &   3 &     \\
     &   9 & 428 &   . &   2 \\
     &  -1 &  18 & 570 &   . \\
     &   . &  -2 &  31 & 713 \\
     &   . &   . &  -3 &  46  \\
     &   . &   . &     &  -6  \\
\end{array}
\right)
 \, ,
\end{equation}
with all elements rounded to 1 unit. Elements smaller than one unit
were left out.
The dots indicate again the diagonal. The higher order harmonics, measured with the coil probe are in the order of some units.
The basis terms of the toroidal circular multipoles 
[Eq.~\eqref{eq:toroidal-circular-tn-ts-cplx}] are scaled with $\epsilon/4$
and  the 
magnitude of term $T_1$ and $T_2$ is still less then 2. For accelerator magnets one can safely assume that 
 all higher order harmonics are well below 10 units. So one can 
conclude that the effect of this matrix can be neglected for all
measured harmonics except the main one 
if an field description accuracy of not better than
$\frac{\eps}{4}\frac{10}{10~000}$ is required. 
The toroidal circular term 
(see Table~\ref{tab:circular-toroidal-basis-terms} and
Eq.~\eqref{eq:toroidal-circular-tn-ts-cplx}) for $m=1$ gives also a
quadrupole and a term caused by Term $T_2$; thus the perturbation term is then exactly
zero for the normal part. The skew part is $4 x/\rref$ (see
Table~\ref{tab:circular-toroidal-basis-terms}),
 but this can be neglected as the skew component is small ($<$ 10 units) and still has
to be multiplied with $\epsilon$. 
Therefore only a quadrupole 
of $\approx 140 \approx 20 \epsilon$~units and a sextupole of $\approx
3$~units has to be added to the set
of cylindric circular multipoles.  
Then the cylindric circular multipole description 
can be used. 

A measurement procedure for obtaining elliptic circular multipoles was
given in section~\ref{sec:elliptic-multipoles-measurements}, with the
measurements performed at different circles: in the centre of the
magnet and  shifted by $\delta x = \pm~30\, mm$. So one can define
\begin{equation}
  \rc^{\pm} = \rc \pm \delta x \qquad \textnormal{and} \qquad \epsilon^{\pm} = \frac{\rref}{\rc \pm \delta x}.
\end{equation}
The different $\epsilon$ are then given by (in units)
\begin{equation}
  \epsilon^{\pm} = 7.6009  \pm 0.0043
\end{equation}
using the values for SIS100 given in
Table~\ref{tab:toroidal-circular-multipoles-coefficients-different-machines}.
The change of $\epsilon$ is at the $7^{th}$ digit and is thus
significantly smaller than the measurement accuracy obtainable with
the systems given here.
This result shows that the cylindric elliptic multipoles can be used
to treat the measurements of the curved dipole magnets of SIS100. 

The calculations for SIS300 yield a matrix  with numerical values of
\begin{equation}
C^{-1}_{SIS300} = I - \frac{1}{10 000}
\left(
\begin{array}{rrrrr}
   . &   3 &     &     &     \\
 125 &   . &   2 &     &     \\
   3 & 249 &   . &   2 &     \\
     &   9 & 374 &   . &   2 \\
     &     &  19 & 499 &   . \\
     &     &  -1 &  31 & 624 \\
     &     &     &  -2 &  47 \\
     &     &     &     &  -4 \\
\end{array}
\right)
\end{equation}
thus the effect of curvature can be neglected for a coil probe length
of  $600~mm$, if the quadrupole is recalculated.
The SIS300 magnets have a round aperture; thus different
 coil positions do not need to be evaluated. 

The inverse of matrix  $C_{NICA}$ \cite{rupac:nica_magnets,nica:asc10} is given by  
\begin{equation}
\small
C_{NICA}^{-1} = I + \frac{1}{10 000} 
\left(
\begin{array}{rrrrr}
  -1 &   13 &      &      &      \\
 500 &   -2 &   10 &      &      \\
 -25 & 1000 &   -2 &    9 &      \\
   1 &  -75 & 1500 &   -3 &    8 \\
     &    4 & -150 & 2001 &   -4 \\
     &      &    9 & -250 & 2501 \\
     &      &      &   19 & -375 \\
     &      &      &    1 &   33 \\
\end{array}
\right)
\,,
\end{equation}
which shows that the effects roughly increase with $l/\rc'$. 
Using the same $\delta x$, but $\rc' = 15\,m$ one gets 
\begin{equation}
  {\epsilon_{NICA}}^{\pm} \approx 26.67 \pm 0.053,
\end{equation}
with $\epsilon_{NICA}$ in units. The values of the matrix $C_{NICA}^{-1}$
are roughly three times higher than for  $C^{-1}$ (SIS100). Similarly 
$\epsilon_{NICA}$ is an order bigger than $\epsilon$ for SIS100. These
influence will have to be evaluated and compared to the required field 
quality descriptions to see if the evaluation using circular
multipoles is still precise enough.

\section{Conclusions}

In this paper we presented 4 different set of multipoles: cylindric circular multipoles, cylindric elliptic
multipoles, toroidal circular multipoles and toroidal elliptic multipoles. While the 
first is common practice, the others are to be considered advanced.  The cylindric circular and elliptic
multipoles are exact solutions of the potential equation while the later ones are approximative 
ones. The elliptic ones give a concise description within the beam vaccum chamber while the 
toridal ones are approximation of the first order. These approximative ones are 
more straightforward to handle and to interpret than their alternative the global toroidal
coordinates.

Measurement methods, based on rotating coil probes, have been theoretically investigated. 
The validity of the cylindric elliptic ones has been demonstrated on the SIS100 FoS dipole magnet 
comparing the field representation to mapper data.  

%

\end{document}